\documentclass[11pt,hidelinks,reprint,amsmath,amssymb,twocolumn,superscriptaddress,showkeys,aps,pre]{revtex4-2}
\usepackage{graphicx}
\graphicspath{{/}}
\usepackage{dcolumn}
\usepackage{bm}
\usepackage{epstopdf}
\usepackage[format=hang, caption=false, justification=centerlast]{subfig}
\usepackage{multirow}
\usepackage{xcolor}
\usepackage{hyperref}
 
\begin{document}

\title{{Deep learning of multi-resolution X-Ray micro-CT images for multi-scale modelling}}

\author{Samuel J. Jackson}
\email[Corresponding author email: ]{samuel.jackson@csiro.au}
\affiliation{CSIRO Energy, Private Bag 10, Clayton South, Victoria 3169, Australia}

 \author{Yufu Niu}
 \affiliation{School of Minerals and Energy Resources Engineering, University of New South Wales, Sydney, New South Wales 2052, Australia}
 
\author{Sojwal Manoorkar}
\affiliation{Department of Earth Science \& Engineering, Imperial College London, London, SW72BP, UK}
 
 \author{Peyman Mostaghimi}
 \affiliation{School of Minerals and Energy Resources Engineering, University of New South Wales, Sydney, New South Wales 2052, Australia}

 \author{Ryan T. Armstrong}
\affiliation{School of Minerals and Energy Resources Engineering, University of New South Wales, Sydney, New South Wales 2052, Australia}

\date{\today}

\begin{abstract}

Field-of-view and resolution trade-offs in X-Ray micro-computed tomography (micro-CT) imaging limit the characterization, analysis and model development of multi-scale porous systems. To this end, we developed an applied methodology utilising deep learning to enhance low resolution images over large sample sizes and create multi-scale models capable of accurately simulating experimental fluid dynamics from the pore (microns) to continuum (centimetres) scale. We develop a 3D Enhanced Deep Super Resolution (EDSR) convolutional neural network to create super resolution (SR) images from low resolution images, which alleviates common micro-CT hardware/reconstruction defects in high-resolution (HR) images. When paired with pore-network simulations and parallel computation, we can create large 3D continuum-scale models with spatially varying flow \& material properties. We quantitatively validate the workflow at various scales using direct HR/SR image similarity, pore-scale material/flow simulations and continuum scale multiphase flow experiments (drainage immiscible flow pressures and 3D fluid volume fractions). The SR images and models are comparable to the HR ground truth, and generally accurate to within experimental uncertainty at the continuum scale across a range of flow rates. They are found to be significantly more accurate than their LR counterparts, especially in cases where a wide distribution of pore-sizes are encountered. The applied methodology opens up the possibility to image, model and analyse truly multi-scale heterogeneous systems that are otherwise intractable.

\end{abstract}

\maketitle

\section{Introduction}
\label{section_intro}

Multiphase porous materials are ubiquitous in engineered and natural systems, for example in electrochemical applications such as fuel cells or batteries \cite{moussaoui2018} in organic matter such as blood vessels \cite{fantazzini2003} and in geological media such as a sandstone or carbonate rocks \cite{vernon2018}. In these systems, heterogeneities in the porous structure often exist at multiple scales, which can range from micrometers to kilometres in the case of the geological subsurface \cite{Ringrose:2015ri}. The smaller-scale heterogeneities can often have larger-scale macroscopic impacts \cite{menke20184d}, and understanding the interaction between multi-scale heterogeneities is key to predicting large scale phenomena \cite{Jackson2020}. However, methodologies to characterise and model multi-scale heterogeneous systems in a tractable way are lacking. From a materials science point of view, multi-scale heterogeneities are increasingly being used to optimise function, in bio(inspired) materials, structural materials and heterogeneous catalysts \cite{Vasarhelyi2020}. Therefore an applied approach to model flow in multi-scale heterogeneous systems is needed for advancements in the material sciences and geological-based disciplines. 

With the advent of X-ray micro-computed tomography (micro-CT), the micro-structure of porous systems can be imaged and visualised, allowing phase segmentation and direct characterisation. Typically, porous samples of $O($cm$^3)$ can be imaged at a resolution of several micrometres, creating images of size 1000$^3$-3000$^3$ voxels \cite{da2021deep}. Following image segmentation, models at the pore-scale of the image can be built to directly predict, e.g., fluid flow \cite{raeini2012}, electrical resistivity \cite{man2000}, and reactive transport \cite{mostaghimi2016}. The segmented images can also be used directly for materials characterisation and to detect material defects \cite{Vasarhelyi2020}. The resolution of the image is intricately tied to the accuracy, both in a numerical model sense, and in the ability to resolve key small-scale features \cite{menke2021}. The field-of-view (FOV) of the image controls the ability of the model to capture larger-scale features, and therefore its ability in predicting realistic engineering relevant properties at scales of interest, e.g., effective permeability \cite{Mostaghimi2012}. However, there are inherent FOV and resolution trade-offs in traditional absorption-based X-Ray micro-CT imaging - typically a resolution 3 orders of magnitude below the FOV is possible \cite{cnudde2013}. With ever increasing hardware specifications, these trade-offs are continually reducing, but there are technological limits to what is achievable, especially when considering several orders of magnitude separation in scale (which are common in geological media). Alternative approaches are therefore needed to fully understand, characterise and model multi-scale porous system. 

Alternative approaches generally either try to create breakthrough hardware implementations, or develop software/modelling based improvements. Recently, advances have been made in combining X-ray and neutron dark-field imaging using grating interferometry to allow sub-resolution feature size quantification \cite{Blykers2021}. On the software and modelling side, a common approach is to implement some sort of homogenisation \cite{lydzba1998}. In this approach, the physics of small-scale phenomena are captured in an upscaled, continuum manifestation through the form of an averaging law. The small-scale physics are effectively captured in the coarser, large-scale model allowing the prediction of macroscopic features \cite{Jackson2020}. In the digital rock physics space, researchers have recently utilised homogenisation approaches to incorporate sub-resolution micro-porosity into multi-scale models, allowing effective characterisation of transport properties over macroscopic scales, whilst including sub-resolution impacts \cite{ruspini2021, wang2021}. Statistical based models have also shown promise, whereby lower resolution images use well-resolved statistics linked to high-resolution features to model large-scale systems accurately \cite{Botha2016}.   

While these approaches offer improvements over the aforementioned direct methods, and increase the size (and effective resolution) of systems that can be analysed, step change advances are potentially possibly through the application of deep learning (DL) supported by advances in graphics processing unit (GPU) hardware and network architectures \cite{da2021deep}. Since deep learning development has been rooted in computer vision and photographic image processing, many approaches are readily suited to offer improvements for X-ray micro-CT imaging and modelling \cite{da2021deep}. In the computer vision community, super-resolution (SR) is the classical ill-posed problem aiming to reconstruct high-resolution (HR) images from low-resolution (LR) images, essentially trying to directly enhance image resolution beyond hardware limits. With this, coarse images taken across large FOV could be enhanced artificially to the level required to perform accurate modelling, thus circumventing traditional trade-offs in analysing multi-scale porous systems. 

Conventional SR methods include projection onto convex sets (POCS) \cite{Panda2011, Shen2014}, Bayesian analysis \cite{Tipping2003, Pickup2008}, example-based \cite{Freeman2002, Salvador2016} and sparse representation \cite{Yang2008, Yang2010}. However, these conventional SR methods have drawbacks that limit their practical use, for instance, example-based approaches need to learn from large dictionary pairs at the expense of excessive computational times \cite{Zhu2014} and POCS-based approaches usually cause under-constrained solutions \cite{Park2003}. With the rapid development on GPUs, deep learning algorithms have since achieved state-of-the-art performance on SR reconstruction problems. DL-based SR methods aim to learn the hierarchical representation of data by an end-to-end mapping between LR and HR data. Instead of learning the dictionary mapping as per traditional SR approaches \cite{Timofte2013, Zhu2014}, DL-based SR methods learn the information via multiple neural network layers implicitly \cite{Dong2014}. Once trained, the DL networks can be used on unseen data to create SR images readily from LR images, creating large, multi-scale datasets for analysis. Dong et al. (2014) first proposed a super-resolution convolutional neural network called SRCNN based on patch-based data and achieved superior performance compared to traditional SR methods. Since then, many deep neural network architectures have been developed for the SR problem \cite{Dong2016, Kim2016, Shi2016, Lim2017, Ledig2017, Wang2018}. Approaches using paired (i.e. registered) training data generally achieve the most accurate results. However, as paired-data is not always available, unpaired DL models have also been developed based on generative adversarial networks (GANs) \cite{Bulat2018, Yuan2018} which have also shown accurate results. 

There are recent examples of DL-based SR models applied to X-ray micro-CT data for digital rock analysis. Typically, pixel-wise metrics, such as Peak Signal to Noise Ratio (PSNR) and Structural Similarity Index Measure (SSIM) have been the primary means to monitor model performance and assess similarity between SR and HR images \cite{Wang2019, Wang2019b,Chen2020}. These works mainly focused on the SR reconstructed image quality itself. \cite{Wang2020, Niu2020} developed unpaired GAN approaches to improve micro-CT image resolution, validating the results using petrophysical property predictions, such as permeability and porosity, as well as geometrical metrics, such as the Euler characteristic. All of these approaches (apart from \cite{Niu2020}) used LR images created from downsampled HR images, not through direct optical manipulation with the micro-CT hardware. The true level of noise transferral across scales is difficult to directly assess when using images generated from one another, even if noise is added.  Although \cite{Niu2020} did use optically generated HR data for training, their GAN approach was validated on unpaired data. \cite{Janssens2020} were the first to demonstrate a super-resolution method using paired LR and HR data obtained using hardware-based optical magnification. They also compared against artificially downsampled LR data across a number of petrophysical and geometric measures, demonstrating the ability of deep learning methods to produce physically realistic super-resolution images.   

Three critical components lacking from these previous studies are: development and testing of the deep learning network on imperfect high-resolution data (e.g. with detector/reconstruction artefacts), direct experimental validation of the SR image properties (i.e. not just numerical simulation verifications), and the application of the approach to a truly multi-scale system, involving heterogeneity and testing data different than the original training data. In this work, we tackle these components directly, using data from high-quality multiphase flow experiments performed on multiple samples, imaged at varying resolution with some image defects presents. The experimental dataset is from \cite{Jackson2020} and utilises two distinct Bentheimer sandstone cores of diameter 1.25cm, length 6-7cm, with varying cm-scale layered heterogeneities. It consists of 6 $\mu$m resolution images of the whole cores at dry, brine saturated and partially saturated states during steady-state decane-brine fractional flow experiments with continuous pressure measurements. We enhance the dataset by performing additional X-ray imaging of the cores in their dry state at varying resolution using optical magnification, creating paired data in 4 subvolumes at 2,6, and 18 micrometer resolution. 

\begin{figure*}
\includegraphics[width=\textwidth]{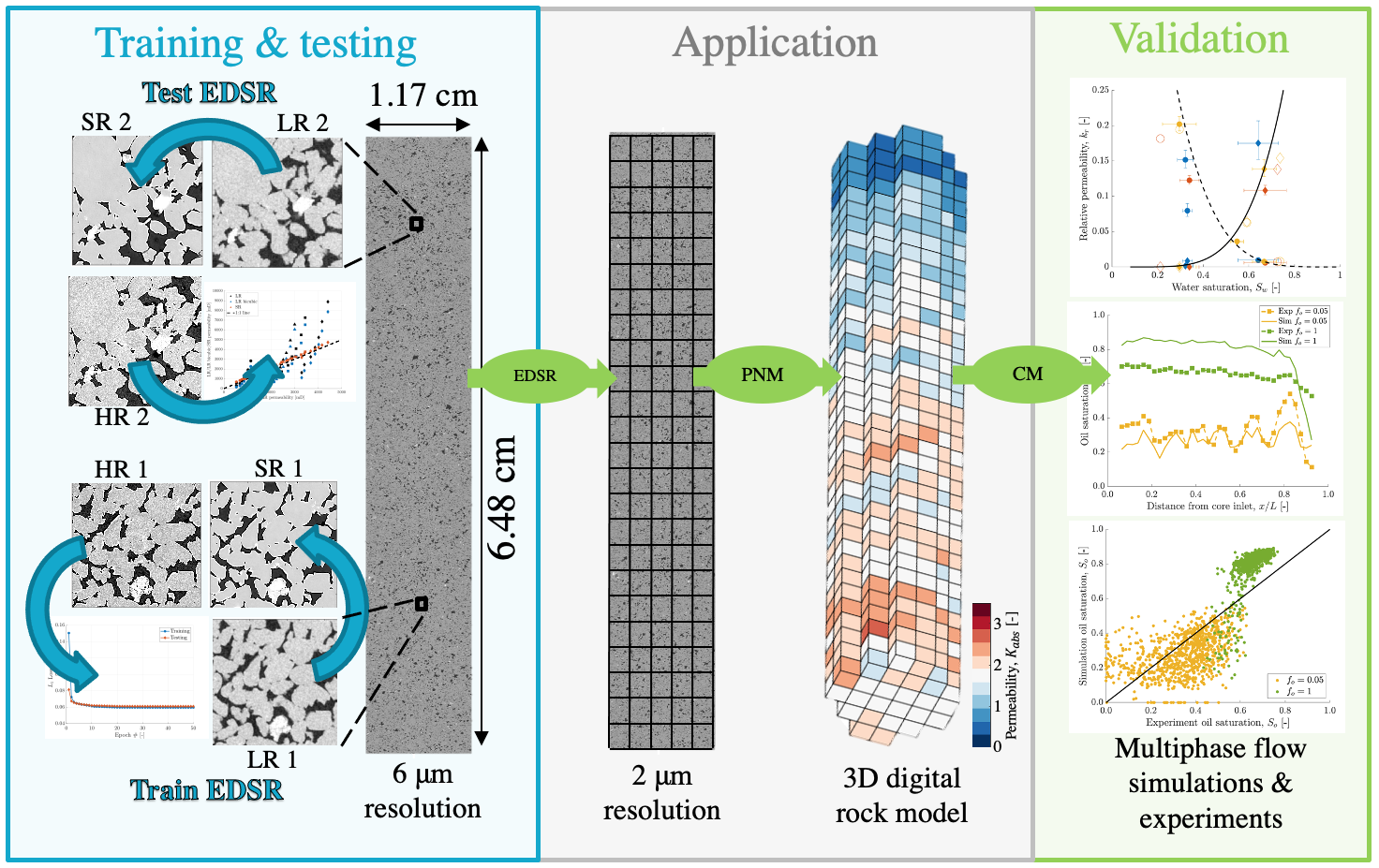}
\caption{Summary of the multi-scale characterisation and modelling workflow, flowing from left to right. First, the EDSR network is \textit{trained \& tested} on paired LR and HR data to produce SR data which emulates the HR data. Second, the trained EDSR is \textit{applied} to the whole core LR data to generate a whole core SR image. A pore-network model (PNM) is then used to generate 3D continuum properties at REV scale from the post-processed image. Finally, the 3D digital model is \textit{validated} through continuum modelling (CM) of the muiltiphase flow experiments. Note, here we show the workflow using core 1. For core 2, we don't perform any extra training - we use the same trained EDSR model and apply this straight to the unseen LR data.}
\label{fig_workflow} 
\end{figure*}

The data is used to develop and train a 3D Enhanced Deep Super Resolution (EDSR) convolutional neural network, based on the original 2D implementation in \cite{Lim2017}. Distinct from previous work, we use a single subvolume to train the network, but test the network on unseen data from other subvolumes in different samples, with regions of varying micro-structure. The unseen data is from a similar rock type (sandstone) as the training set, but has a distinct micro-structure and flow properties. We validate the SR results using conventional image metrics and pore-network model (PNM) flow simulations, comparing to the paired ground-truth HR data.  We do this across multiple image processing realizations, giving an unbiased comparison between results \cite{Garfi2019,leu2014fast}. We then apply the trained EDSR network to generate whole core images of the full samples, allowing us to analyse key multi-scale features that impact the large scale flow. We perform the EDSR SR generation in parallel, on $\approx1000$ distinct volumes of size 1000$^3$ voxels for each core, creating whole core images of total size $\approx 6000 \times 6000 \times 32000$ voxels,  significantly larger than direct imaging would permit. The high-resolution subvolume images (1000$^3$ voxels), are used with the PNM to generate spatially varying petrophysical properties across the samples. We utilise the network modelling approach to alleviate direct modelling computational costs, which would be infeasible for $\approx$1000 distinct volumes of size 1000$^3$ voxels, even on super computing clusters. The petrophysical properties are combined to create 3D continuum models, which we use to simulate the experiments directly. We can then compare measured pressures and 3D saturations at varying scales directly to the experiments, thus providing a true validation of the multi-scale performance of the deep learning method. The modelling approach is predictive, in the sense that continuum properties are directly generated from pore-scale models, meaning the approach is calibration free and deterministic, leaving significantly less ambiguity compared to other approaches. 

The novelty of the work is that we develop a fully tractable and predictive methodology to model multiphase flow in heterogeneous porous systems by coupling deep learning, pore network modelling and micro-CT imaging. The advancements from previous works are threefold; (1) we develop an improved deep-learning method for super-resolving 3D low resolution data, that is applicable to micro-CT images with common hardware/reconstruction defects; (2) we develop an integrated workflow from image acquisition to modelling that allows accurate multi-scale analysis of heterogeneous systems that are otherwise intractable; (3) we verify and validate the approach with numerical and experimental data, in samples of different heterogeneous structure across a range of test conditions. Overall, the presented methodology could be applied to any type of natural or engineered porous system with multi-scale heterogeneity where the optimization of flow and transport is of upmost importance.   

The paper is organised as follows.
\begin{enumerate}
\item  We first present the workflow methodology including image processing, deep-learning, pore-network modelling and continuum modelling. 
\item We then present results comparing various metrics of the generated SR images to the ground-truth HR images. 
\item We finish by using the trained EDSR network to generate SR images of the whole rock samples, which in turn are used to generate continuum models. Continuum model simulations are then compared to experimental results across multiple flow regimes. 
\end{enumerate}

\section{Methodology}
\label{sec_methodology}

\subsection{Overview}
\label{sec_method_overview}

The combined multi-scale characterisation and modelling workflow is summarised in Figure \ref{fig_workflow}. It comprises of image acquisition, training and testing of the deep learning algorithm, followed by application and creation of a 3D continuum model, concluding with continuum-scale multiphase flow simulations and comparison to experiments. 

We demonstrate the methodology on two distinct Bentheimer rock cores -- referred to throughout as core 1 and core 2 -- which have previously undergone extensive experimental and modelling work in \cite{Jackson2020, Zahasky2020}. The cores have diameter, 12.35mm, lengths $73.2$mm and $64.7$mm, core-averaged porosities of 0.203 and 0.223 and permeabilities of 1.636D $\pm$ 0.025D and 0.681D $\pm$ 0.006D for core 1 and 2, respectively. Core 2 has a clear low permeability lamination occurring at  $\approx2/3$ of the total core length, whereas core 1 has a general fining of grains towards the outlet of the core creating a reduction in porosity \cite{Jackson2020}.

The experimental dataset from \cite{Jackson2020} consists of 6 $\mu$m resolution images of the whole cores at dry, brine saturated, and partially saturated states during steady-state decane-brine fractional flow experiments for both drainage and imbibition co-injections. Here, we focus solely on the drainage cycle of the experiments; multi-scale modelling of the imbibition evolution and trapping is the focus of on-going work. We compare to the more complete experiments 2 and 3 in \cite{Jackson2020}, which were performed on core 1 and 2, respectively. The experiments were performed at 1.5MPa pore pressure, at a temperature of 30$^{\circ}$C. Brine and decane (oil) were co-injected at a fixed total flow rate of  0.1 ml.min$^{-1}$. The fractional flow of decane was varied in the drainage cycle from $f_o = 0, 0.05,1$ for core 1 (exp 2), and  $f_o = 0, 0.05,0.5,1$ for core 2 (exp 3).  Each fractional flow was run until steady-state was achieved in the pressure and saturation profiles, generally taking over 1 day. At each steady-state, whole core images were taken. Saturations were reconstructed from mm to cm scale, and core average relative permeabilities were calculated from measured pressure drops across the sample. We use these measures in the final continuum scale modelling and validation. 

We perform extra imaging of the cores to create paired low-resolution (LR) and high-resolution (HR) data in order to train and test the deep learning algorithm. We image two subvolumes for each core, at locations $1/3$rd and $2/3$rds of the way along the core length, at resolutions of 2$\mu$m and 6$\mu$m. The same Zeiss Versa 510 X-Ray micro-CT scanner and protocol is used as in \cite{Jackson2020} -- we use a flat panel detector for the 6$\mu$m, and a 4x optical microscope objective for the 2$\mu$m image. Precise energies, projections, and binning information can be found in the supporting information of \cite{Jackson2020}.

\subsection{Image processing}
\label{sec_image_pro}

The acquired LR and HR tomographic images are first reconstructed using Zeiss reconstruction software, correcting for beam-hardening artefacts, and centre-shift of the sample. The corresponding 2$\mu$m and 6$\mu$m images are then registered using normalised mutual information, and cropped to remove edge artefacts. Final image sizes are 225$^3$ and 675$^3$ voxels for the 6$\mu$m and 2$\mu$m resolution scans, respectively. The images are then normalised to have consistent grey scale values \cite{Jackson2020}:
\begin{equation}
CT_{new} = S(CT_{old} + O),
\end{equation}
where the scaling factor, $S$ and the offset $O$ are given by:
\begin{equation}
S = \frac{CT_{ref2} - CT_{ref1}}{CT_{pk2} - CT_{pk1}}, \;\;\;\;    O = \frac{CT_{ref1}}{S} - CT_{pk1}.
\end{equation}

Here, $CT_{ref1}$ and $CT_{ref2}$ are the reference values for the pore and grain space (chosen as 2000 and 10000, respectively) and $CT_{pk1}$ and $CT_{pk2}$ are the measured peak values for the minimum and maximum greyscale phases present in the image. $CT_{pk1}$ and $CT_{pk2}$ are found from the grey-scale histogram using a MATLAB in-house peak detection algorithm. We then convert the images from 16 bits to 8 bits:
\begin{equation}
CT_{8bit}  = \frac{CT_{16bit} - CT_{min} }{ CT_{max} - CT_{min}},
\end{equation}
where $CT_{16bit}$ is the input voxel value, $CT_{min}$ and $CT_{max}$ are the related extremes determined from the grey-scale histograms, here taken as 0 and 2$\times$10$^4$, respectively. At the end of this processing, we have 2 subvolumes of normalised, registered LR and HR data for core 1 and core 2, giving 4 subvolumes in total at two resolutions. We note that we also obtained very low resolution data (VLR) at 18$\mu$m for the subvolumes, using the same workflow. This was not used here, since the whole core images were at 6$\mu$m, but is included in the data share (see Section \ref{sec_data_access}) for further work. 

\subsection{Deep learning}

\begin{figure*}
\subfloat[]{\includegraphics[width=0.9\textwidth]{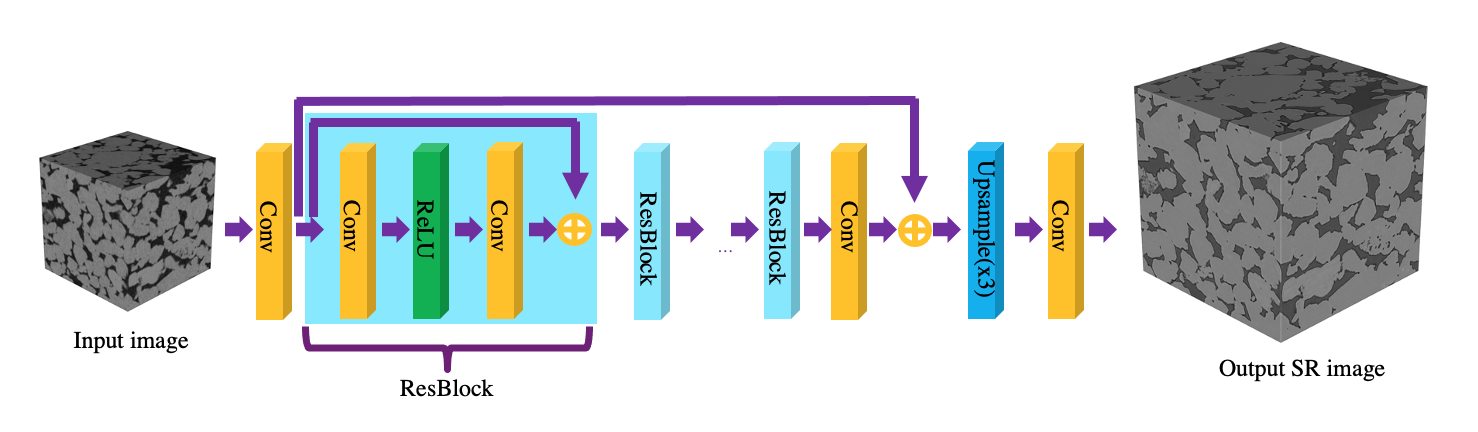}}

\subfloat[]{\includegraphics[width=0.4\textwidth]{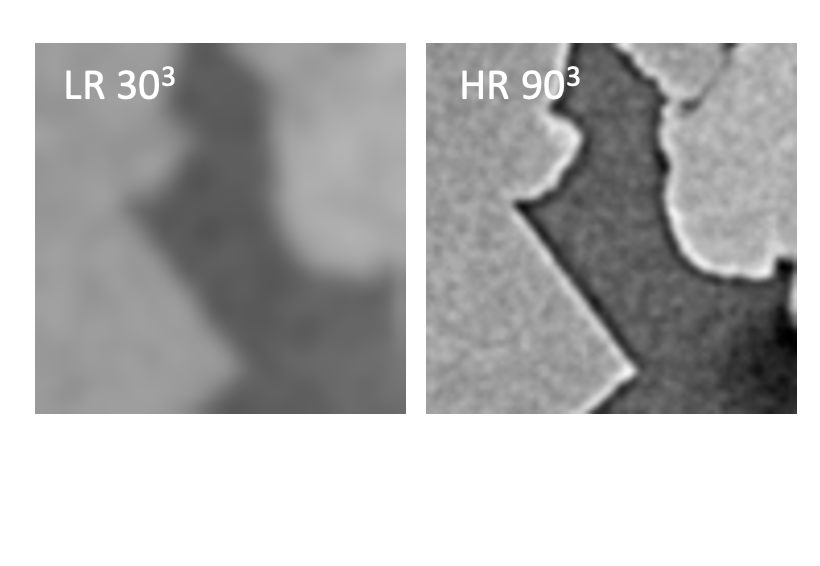}}\quad
\subfloat[]{\includegraphics[width=0.4\textwidth]{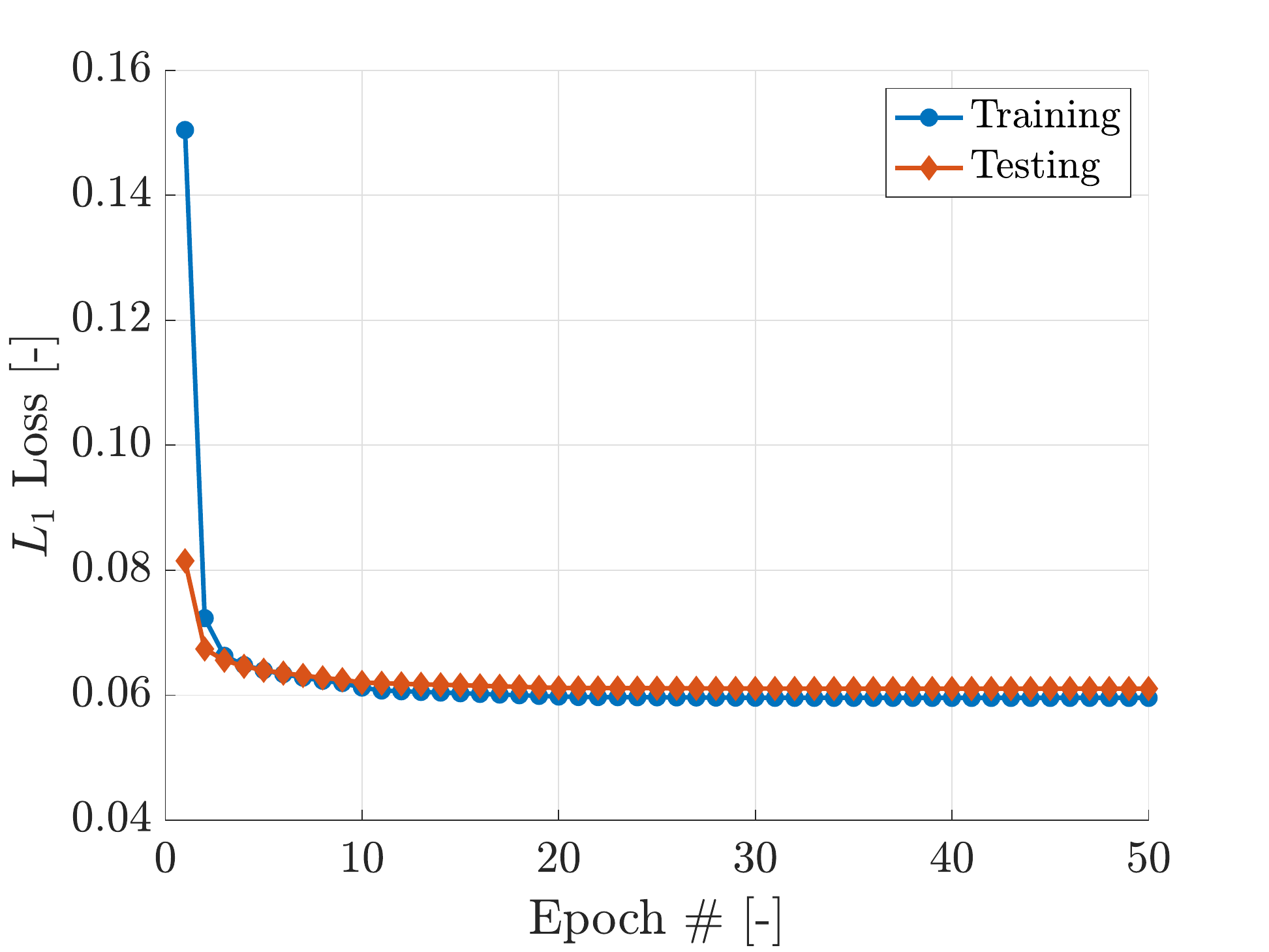}}
\caption{Deep learning workflow and testing/training. (a) EDSR network architecture. (b) Example testing images, LR image is at 6 micron resolution, HR image is at 2 micron resolution. (c) $L_1$ loss function for the EDSR network training and testing over 50 epochs.}
\label{fig_deep_learning_workflow} 
\end{figure*}

To train the deep learning algorithm, we only use the paired data from core 1, subvolume 1, leaving the other 3 subvolumes for validation. Image patches are extracted from the LR and HR data with size 30$^3$ and 90$^3$ voxels, respectively, through an overlapping moving window (15 and 45 voxel overlap for each patch, respectively). This creates 8000 patches, from which we use 6400 patches for training and 1600 patches for validation in a 80/20 Pareto split. Figure \ref{fig_deep_learning_workflow}b shows a 2D slice of a typical training image patch at low and high resolution. The patch size was chosen to cover the largest typical pores present in the Bentheimer samples of $O(100 \mu$m).

We employ a 3D EDSR model as our deep learning algorithm, shown in Figure \ref{fig_deep_learning_workflow}a. The EDSR encompasses a sequence of convolutional layers, residual blocks \cite{He2015}, and an upsampling module. Along with skip connections in the residual blocks, the EDSR can alleviate gradient vanishing/exploding problems. We use a modified version of the original 2D EDSR structure from \cite{Lim2017}. Instead of using pixel shuffle upsampling methods proposed by \cite{Shi2016}, we utilise trilinear upsampling in our 3D implementation. In addition, we also reduce the numbers of residual blocks and filters (12 residual blocks and 24 filters in each convolutional layer) appropriately to improve computational efficiency and in an effort to reduce over-fitting of artefacts apparent in high-resolution micro-CT image, specifically ring and beam hardening artefacts. These are discussed in more detail in Section \ref{sec_results_pore_scale}. 

We utilise the Adam optimisation algorithm for training \cite{Kingma2014}, with a batch size of 32 and total epochs of 50, chosen through user experience in the training process. The learning rate is initially set to 1 x 10$^{-5}$ and decreased tenfold every ten epochs to consolidate the training process. The $L1$ loss metric is utilised to train and monitor the EDSR network:
\begin{equation}
L_{1, loss}  = \sum_{i = 1}^n \left| CT_{i,HR} - CT_{i,SR}, \right|
\end{equation}
where $CT_{i,HR}$ and $CT_{i,SR}$ are the pixel greyscale CT values for the HR and the generated super resolution (SR) images, respectively. The training and testing $L_{1, loss} $ is shown in Figure \ref{fig_deep_learning_workflow}c, highlighting convergence to a low loss (3x reduction from the initial) after $\approx$10 epochs. The training was implemented in Pytorch using a NVIDIA GeForce RTX 2080Ti. 

Once trained, the EDSR network can be utilised to generate SR images with a 3$\times$ increase in resolution from any input LR images (for other integer factors, eg 2$\times$, 4$\times$, a new network has to be trained on corresponding data). We utilise the EDSR to generate corresponding SR images for the unseen LR images from core 1 subvolume 2, and for core 2 subvolume 1 and 2. With corresponding HR data, these images can be directly evaluated to assess the efficacy of the EDSR network in producing an SR image which emulates the HR image. As a further comparison, we also generate a high-resolution image using a simple 3D cubic interpolation of the LR image to the required 3$\times$ resolution, with the result labelled `Cubic'. This allows comparison with common polynomial interpolation methods that can be used to increase the resolution of an image without using any prior knowledge of the image or trained weights. This also acts as example of what a simple resolution enhancement can achieve when performing network extractions and flow simulations which are often mesh dependent.
\begin{figure}
\subfloat[]{\includegraphics[width=0.45\textwidth]{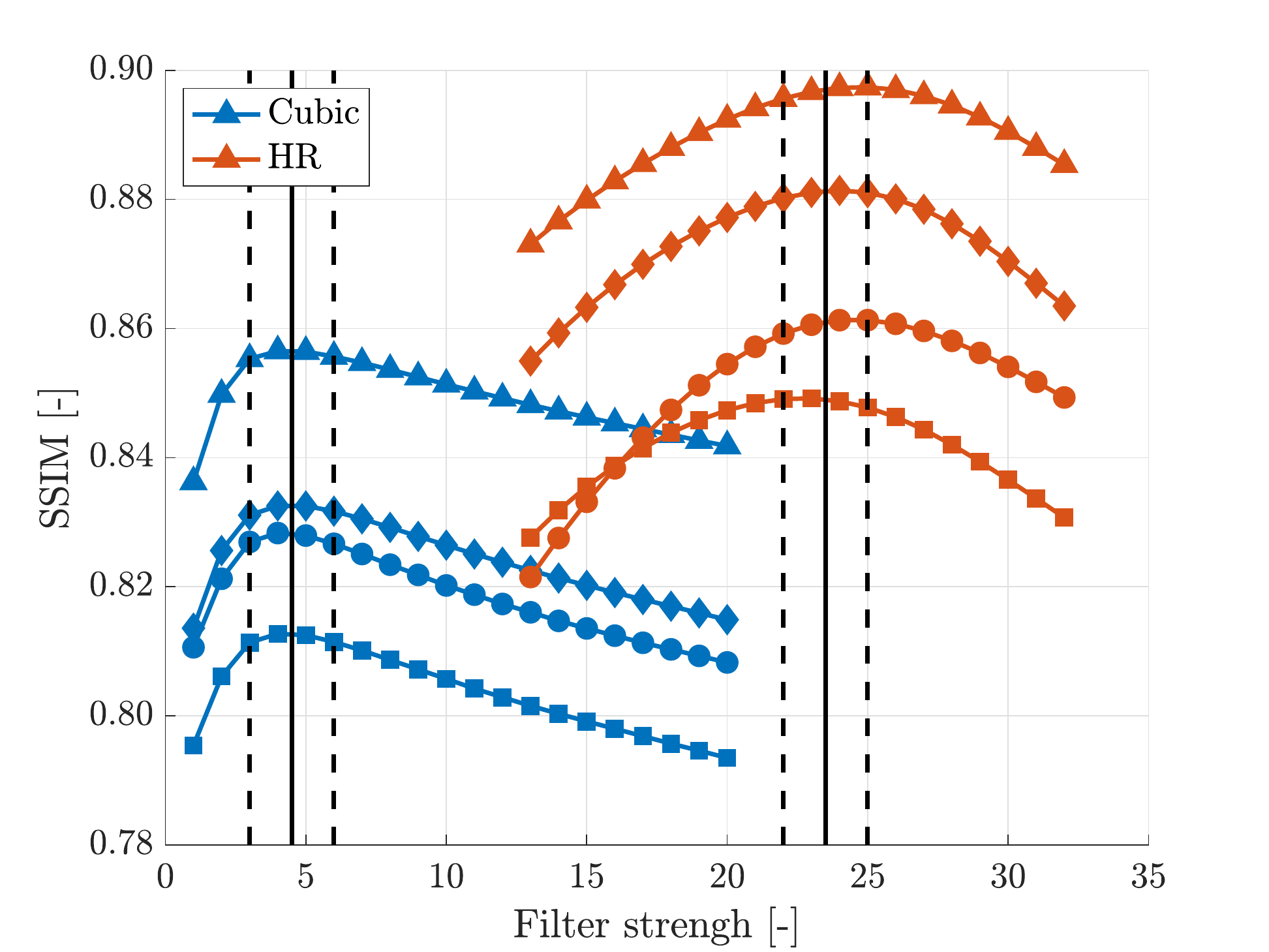}}

\subfloat[]{\includegraphics[width=0.45\textwidth]{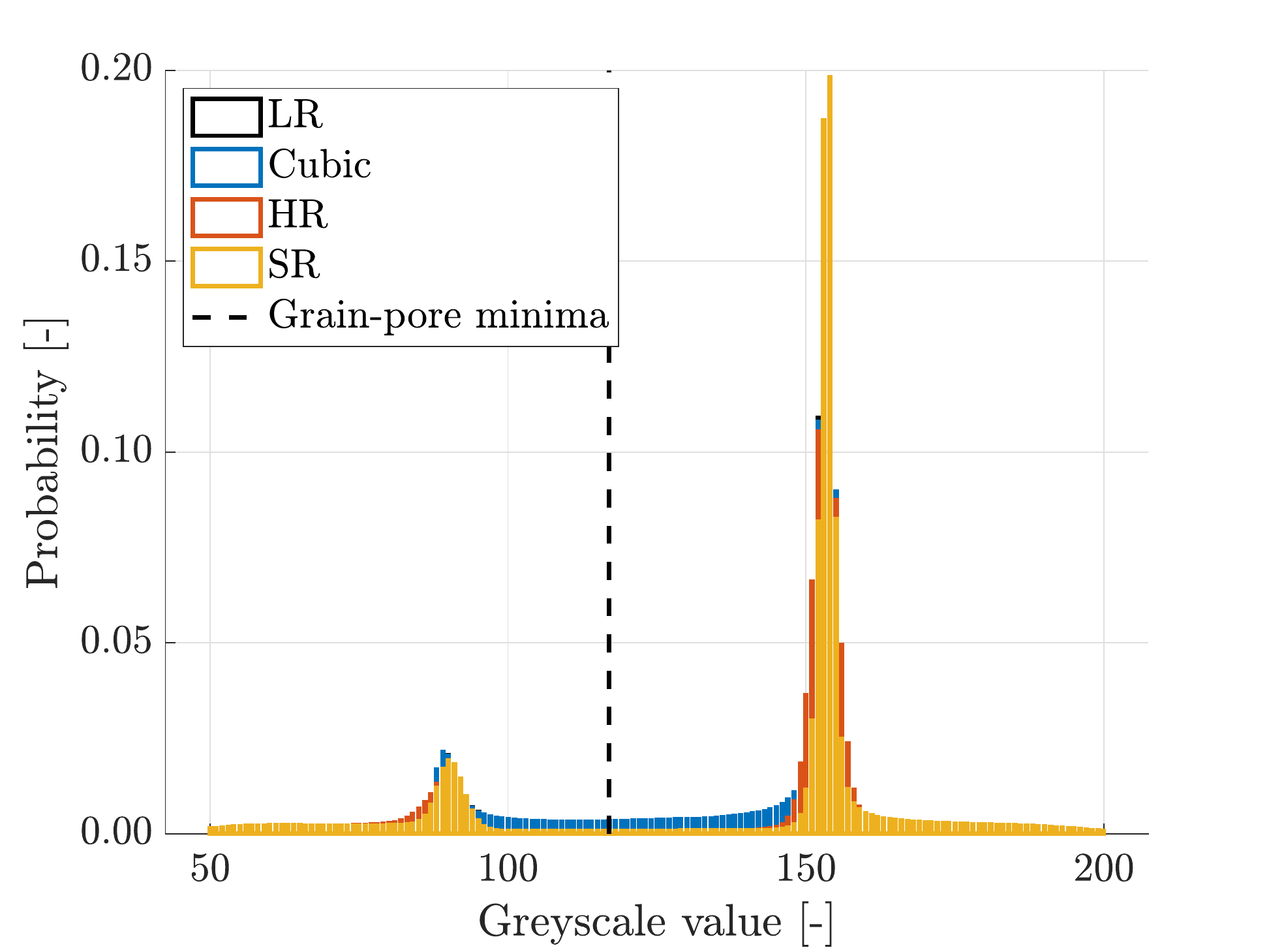}}
\caption{Image processing steps to allow quantitative segmentation and flow comparisons. (a) Structural similarity index measure (SSIM) of the filtered Cubic and HR image compared to the SR image. Different symbols refer to the subvolumes. Traingles - core 1, subvolume 1. Diamonds - core 1 subvolume 2. Squares - core 2 subvolume 1. Circles - core 2 subvolume 2. The optimum filter strength is shown as the solid black lines for the Cubic and HR images (dashed lines highlight the maximum). (b) Histograms of greyscale values for the filtered core 1 subvolume 1 image, showing base segmentation threshold (minima between grains and pore  = 117 greyscale value) and overlapping distributions of the normalised images.}
\label{fig_SSIM} 
\end{figure}

To quantitatively compare the paired LR, Cubic, SR, and HR images across the samples, we use a mix of image analysis and flow simulations performed with pore-network modelling. To this end, the images must be segmented into binary pore and grain phases, so that pore-networks can be extracted. 

Due to the differing noise levels in the images, it is useful to filter the images to a common level, such that subsequent segmentations are comparable. It is well known that filtering and segmentation operations in image processing can be highly user specific and biased \cite{Garfi2019}, if objective measures of the output are not used.  We perform the filtering operation objectively using the structural similar index measure ($SSIM$) \cite{Wang2004}. Since the SR image generated from the EDSR network is essentially filtered in the ResBlocks (see Figure \ref{fig_LR_HR_SR_comps} below), we filter the LR, Cubic, and HR images to a similar level using the SR image as a reference. To compare and optimise the resultant output to have similar filtering levels, we use the $SSIM$ between the images and the SR image, given by:
\begin{equation}
\label{eq_SSIM}
SSIM(x,y) = \frac{ \left( 2 \mu_x \mu_y + C_1 \right)   \left( 2 \sigma_{xy} + C_2 \right)}{\left( \mu_x^2 + \mu_y^2 + C_1 \right) \left( \sigma_{x}^2 + \sigma_y^2 + C_2 \right)},  
\end{equation}
where $\mu_x, \mu_y, \sigma_x, \sigma_y$, and $\sigma_{xy}$ are the mean, standard deviation, and covariance between images $x, y$. $C_1$ and $C_2$ are regularization constrains for regions with mean or standard deviation close to zero, for the images here they are given by $(0.01d)^2$ and $(0.03d)^2$, where $d$ is equal to the dynamic range of 255. To filter the LR, Cubic, and HR images, we use  a non-local means filter with varying filter strength \cite{buades2007}. The search window and comparison window size are set to maximise the SSIM with 11 and 5 voxels, respectively, and increased 3$\times$ for the high-resolution images. 

In Figure \ref{fig_SSIM}a we compare the SSIM for the Cubic and HR image to the SR image for varying filter strengths, across all four subvolumes. We cannot directly compare the LR image due to the image size mismatch with the SR image. None the less, we see that the HR image shows a greater similarity with the SR image across all filter strengths than the corresponding Cubic image. In the figure, we show the optimum filter strength for both cases, which is chosen to maximise the SSIM across all subvolumes. We use these filtering parameters for all images in subsequent processing.

Previous studies comparing image enhancement methods have typically used a single user defined segmentation threshold, based on for example ``optimal balance between over- and underestimating connectivity" \cite{Janssens2020}. This makes objective comparison between outputs difficult. Here, we use a range of threshold values, in the spirit of \cite{Bultreys2018, Zahasky2020,leu2014fast}, and compare results quantitatively across the range, to indicate the sensitivity and absolute variations. We use a simple threshold segmentation method, with the base threshold level chosen as the minima between the pore and grain peaks in the grey-scale histograms, see Figure \ref{fig_SSIM}b \cite{Zahasky2020}. This minima (value 117) is selected as the average minima for the $LR$ and $HR$ training image, and is close to the minima for all images (SR also, see Figure \ref{fig_SSIM}b). We segment the pore-space based on this threshold, and vary it from -15\% to +15\% in 5\% increments, creating a total of 7 segmentations for each subvolume.

\subsection{Pore-network modelling}

The segmented LR, Cubic, HR, and SR images are used in a pore-network modelling workflow to predict key petrophysical parameters. Pore-networks are extracted from the binary images using the conventional network extraction tool from pnextract \cite{Raeini2017}, an updated version of that was first presented in \cite{Dong2009}. A distance map of each voxel to its nearest solid voxel is used to construct a medial surface of the pore-space with a maximal sphere hierarchy; each sphere has a radius equal to the distance map. An iterative scheme then removes overlapping nested maximal spheres along the medial surface. This results in a set of voxels on the medial surface that have a maximal sphere and connectivity with neighbouring points, representing the pore-space morphology. A watershed segmentation of the distance map is then used to assign each voxel (and neighbours) on the medial surface to a unique pore-body, with the joining faces between bodies defining the connecting throats. 

The resulting pore-network is used in flow simulations using the pnflow package from \cite{Raeini2018}, an updated version of the original \cite{Valvatne2004} algorithm. The conventional network extraction and flow simulation has been validated in previous works \cite{Bultreys2018, Zahasky2020}, as well as the more complex generalised network modelling approach \cite{Raeini2019}. For the flow modelling here, pore-throats are assigned a shape factor based on the underlying pore-space geometry:
\begin{equation}
G = \frac{R^2}{4A},
\end{equation}
where $R$ is the inscribed radius of the maximal sphere and $A$ is the throat's cross sectional area \cite{Bultreys2018}. Pore-bodies use a shape factor average from the connected throats, weighted by the connected cross-sectional area of the throats. Based on $G$, the network elements have either triangular, square, or circular cross-sections, dependent on the following criteria \cite{Valvatne2004, Raeini2017, Raeini2018}:
\begin{align}
0 \le &G \le \frac{\sqrt{3} }{36},\;\; \;\; \text{triangular} \\
\frac{ \sqrt{3} }{36} <  &G \le 0.07, \;\;\; \text{square} \\
0.07  <  &G.\;\; \; \; \;\;\; \; \; \;\;\; \; \;  \text{circular }
\end{align}
Quasi-static capillary dominated drainage flow is simulated in the network elements using fluid interface force balances with the Mayer-Stowe-Princen method \cite{Mason1991}. All throats at the network inlet are assumed filled with non-wetting phase. In each displacement step, a wetting and non-wetting phase pressure are prescribed to give a defined capillary pressure, $P_c = P_{nw}- P_{w}$. Throats and pores are then drained in order of increasing capillary entry pressure controlled by the network element shapes and contact angles. In the strongly-water systems considered here, we assume a contact angle of 45 $^{\circ}$C to be comparable to mercury intrusion experiments and the drainage flow experiments \cite{Jackson2020}. Arc menisci can form due to small wetting layers in the polygonal elements, which require complex entry pressure calculations, given by \cite{Mason1991, Valvatne2004}.  Once all available and accessible elements have been filled, the network saturation can be found based on the invaded network volume, giving a macroscopic $P_c(S_w)$. Incrementing the non-wetting phase pressure in steps until the irreducible saturation simulates a full drainage cycle. 

At each equilibrium stage, connected-pathway single/multiphase transport can be simulated through the network. Both hydraulic (absolute \& relative permeability) and electrical (resistivity, formation factor) properties are calculated. Conservation of flux (electrical or mass) $Q$ is imposed at each pore-body, $p$, through adjoining throats, $t$:
\begin{equation}
\sum_{t \in p} Q_t ^{\alpha} = 0,
\label{eqn_flux_cons}
\end{equation}
where $\alpha$ is the fluid phase. The local flux between adjoining pores $i$ and $j$ is given by:
\begin{equation}
Q_{ij}^\alpha  = \frac{g^{\alpha}_{ij} }{L_{ij} } \left( \Phi_i^{\alpha} -  \Phi_j^{\alpha},  \right)
\label{eqn_conduc}
\end{equation}
where $g^{\alpha}_{ij}$ is the conductance for phase $\alpha$ between pores of center-separation $L$, and $\Phi$ is the potential. Equations (\ref{eqn_flux_cons}) \& (\ref{eqn_conduc}) can represent both fluid or electrical flow, changing the potential and phases accordingly. The conductance is the harmonic mean of the two pore-body conductances and the connecting throat element. The hydraulic conductance for a single element is calculated analytically using Poiseuille's law for single-phase flow, based on the element geometry \cite{Valvatne2004}.  For multiphase flow, corner and layer flows can occur - here we use empirical conductances based on the flow geometry, validated with laminar Stokes flow simulations \cite{Valvatne2004}. For electrical conductivity, we only consider the water phase conductivity at fully saturated conditions. The electrical conductance for an element is given by the bulk water conductance multiplied by the element cross sectional area. 

A linear set of equations for pore-body potentials can be created from Equations (\ref{eqn_flux_cons}) \& (\ref{eqn_conduc}), with appropriate potential boundary conditions at the inlet and outlet faces (constant potential difference of 1 Pa (pressure) or 1 volt (electrical)), and zero flux boundaries on the other faces. The system can be formed for single phase flow, multiphase connected pathway flow, and electrical flow.  Solving this system, the inlet and outlet fluxes can be reconstructed by summing across the appropriate elements. With this, the absolute permeability, relative permeability, and formation factors can be calculated:
\begin{align}
K &= \frac{ \mu_w Q_{sw} L}{ A \left( \Phi_{inlet} - \Phi_{outlet} \right)},   \\
\label{eqn_perms} k_{r,\alpha} &= \frac{ Q_{\alpha} } { Q_{sw} },  \\ 
F &= \frac{ Q^e_{sw} L}{ \sigma_w A \left( \Phi^e_{inlet} - \Phi^e_{outlet} \right)}  ,
\end{align}
 where $\mu_w$ is the water viscosity, $Q_{sw}$ is the fully saturated water flux,  $Q_{\alpha}$ is the phase fluid flux at intermediate saturation, $Q^e_{sw}$ is the electrical flux in the water saturated system, $\Phi_{inlet} - \Phi_{outlet}$ is the pressure drop across the system, $\Phi^e_{inlet} - \Phi^e_{outlet}$ is the electrical potential drop across the system, and $\sigma_w$ is the bulk water electrical conductivity.
 
We use bulk fluid properties equivalent to those in the brine-decane flow experiments described previously. Fluid viscosities are 7.83$\times$10$^{-4}$Pa.s \cite{Goldsack1977} and 8.03$\times$10$^{-4}$Pa.s \cite{Linstrom1997}, and densities are 1023.2kg.m$^{-3}$ \cite{Ghafri2012} and 723.8kg.m$^{-3}$ \cite{Linstrom1997}, for the brine and decane, respectively. We use an equilibrium contact angle of 45$^{\circ}$ and interfacial tension of 51 mN/m. We use a residual saturation of brine, interpreted from the experiments as $S_{wirr} = 0.08$ and an electrical resistivity of brine as 1.2 $\Omega \cdot$ m.

The petrophysical properties calculated using the pore-network model are used for verification of the deep-learning image enhancement. Results are compared across the different segmentations for each LR, Cubic, HR and SR subvolume, allowing for direct quantitative verification. 

\subsection{Continuum modelling}
\label{sec_methods_cont_model}

Following verification of the deep learning enhancement (e.g. training, testing, and PNM verification), we further validate the approach by performing continuum scale simulations of the aforementioned experiments. To do so, we construct digital models of the whole rock cores, populated with petrophysical properties that vary spatially at the representative elementary volume scale. 

The LR whole core images of core 1 and 2 are first subdivided into representative elementary volumes. The size of these volumes was extensively researched in \cite{Jackson2020}, generally showing cubic side lengths of $>$1.5mm were required to reach representative volumes, where parameter fluctuations dropped to with 5\% of the large-scale variance. We choose the same LR subvolume size as per \cite{Jackson2020, Zahasky2020} -- (316x316x300) and (316x316x318) voxels for core 1 and 2 respectively, ensuring $>$1.5mm in side length.  Following this, the EDSR network detailed previously is used to enhance the LR subvolume images, increasing resolution by 3x. This creates SR subvolumes of size (948x948x900) and (948x948x954) voxels at 2$\mu$m resolution, for core 1 and 2 respectively. 

\begin{figure}
\subfloat[]{\includegraphics[width=0.45\textwidth]{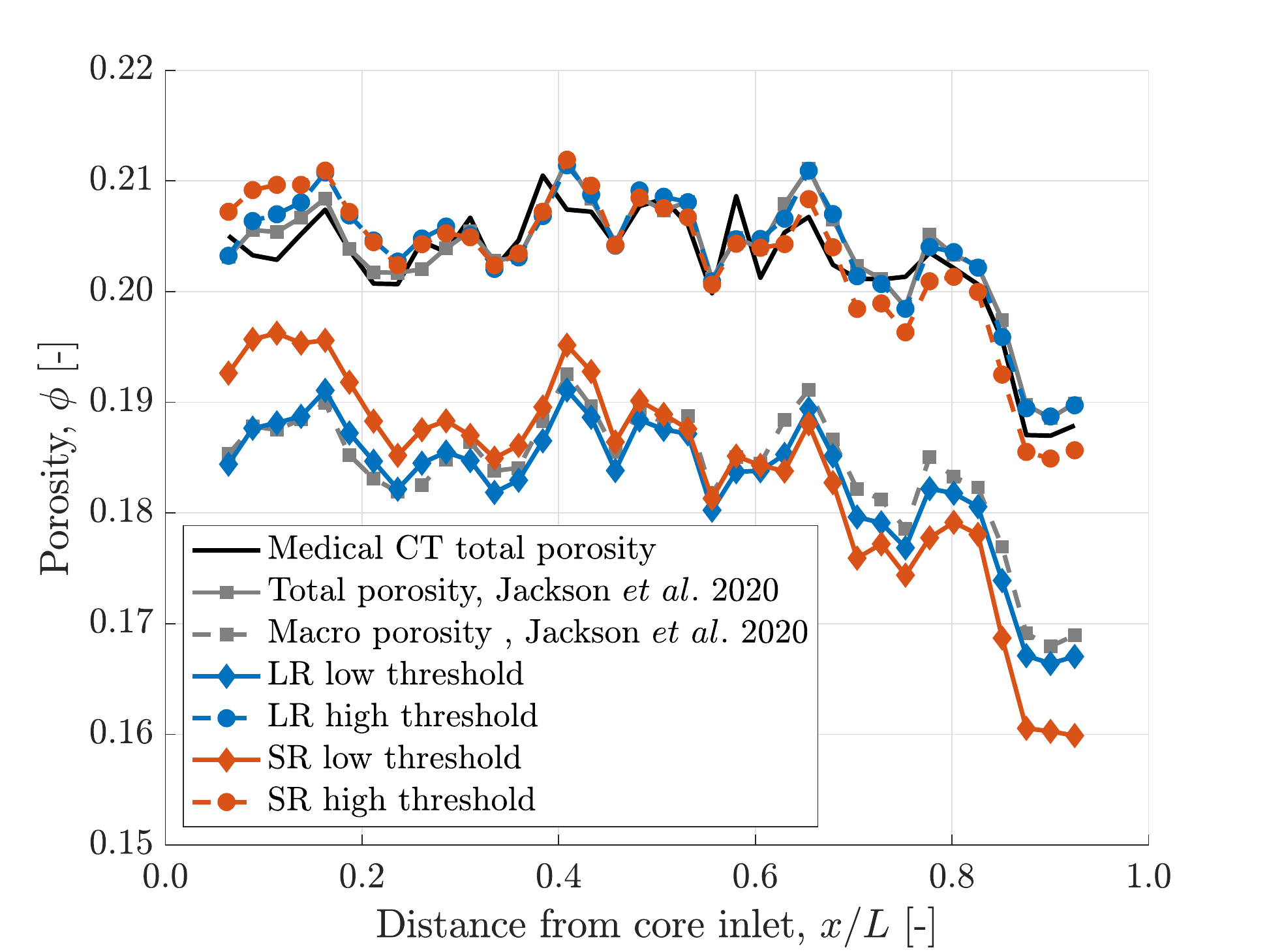}}

\subfloat[]{\includegraphics[width=0.45\textwidth]{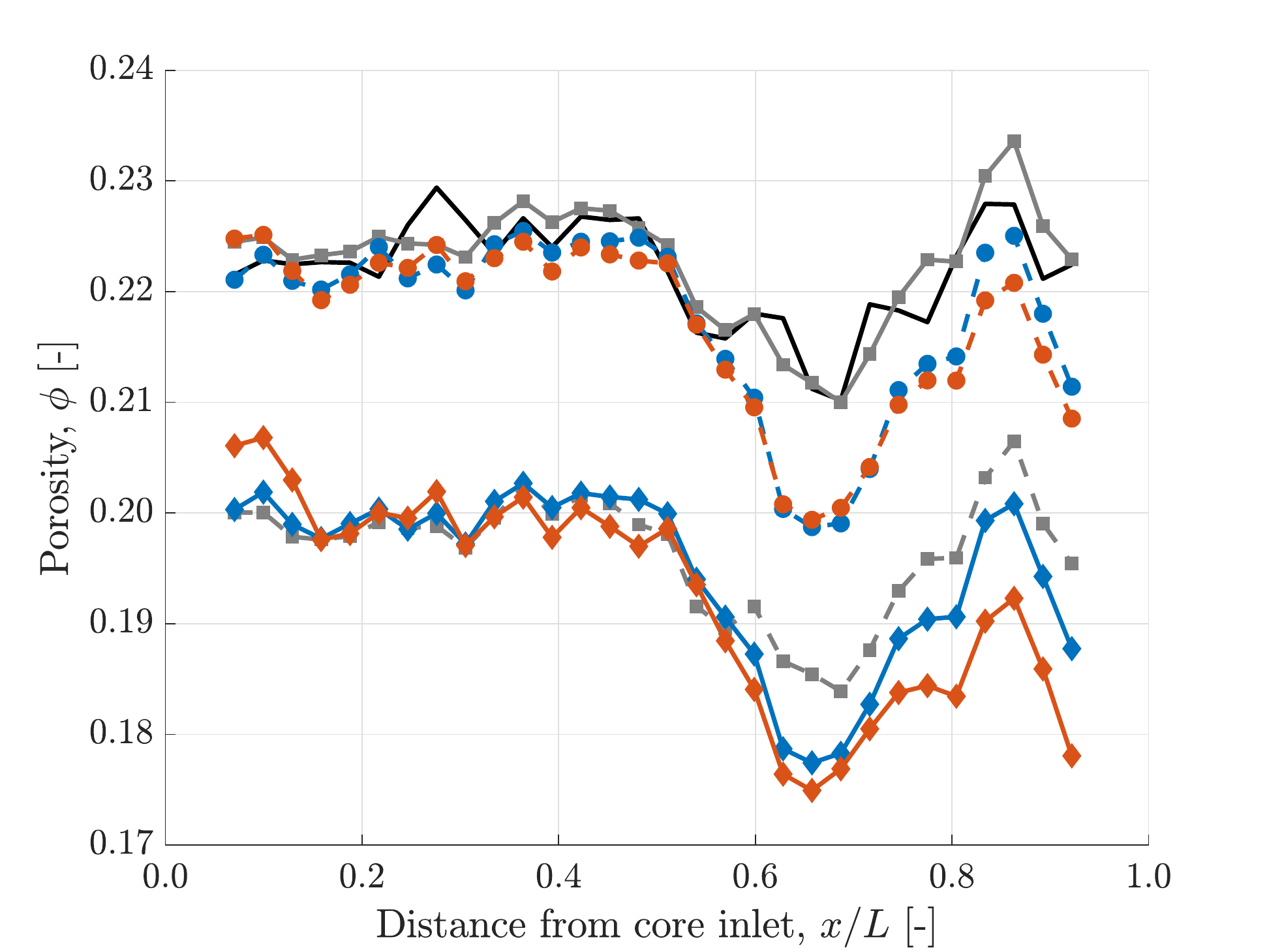}}
\caption{Porosity profiles for core 1 (a) and core 2 (b) for the LR and SR images at two different segmentation thresholds - high threshold (HT) and low threshold (LT) chosen to match the medical CT derived total porosity, and the macro-porosity from earlier analyses \cite{Jackson2020}, respectively. The specific LT/HT threshold segmentation values are: core 1 LR 117/122, core 1 SR 104/117, core 2 LR 118/123, core 2 SR 104/119. }
\label{fig_PNM_porosity_whole_core}
\end{figure}

The LR and SR representations of the whole core images are next filtered and segmented, in a similar vein to the imaging processing section \ref{sec_image_pro}. The filtering protocol is identical, but we segment based on calibration with an externally measured total porosity, derived through medical X-ray CT imaging. The threshold value from the LR and SR images is chosen so that the segmented porosity matches that of the medical CT derived value, shown in Figure \ref{fig_PNM_porosity_whole_core}. The high threshold (HT) value shows the corresponding match for the LR and SR images to the medical CT value, along with a previous `total porosity' segmentation from \cite{Jackson2020}. The slice average porosities from all four images match well. Small discrepancies between the micro-CT segmented images are due to feature representation, and the segmentation choice in \cite{Jackson2020} which used a watershed segmentation. 

As there is likely over segmented pore-space in each image (due to the resolution limit and matching a 'total porosity' measured from medical CT imagery), we perform a further segmentation at a lower threshold (LT). This threshold is chosen to approximate the macro-porosity in the core from \cite{Jackson2020}. This lower threshold removes intermediate grey regions from the pore-space, e.g., clay regions or partial volume effects at pore-grain boundaries. It can give a more representative pore-space for low capillary number flow simulations, since the non-wetting fluid does not drain clay bound, micro-porous regions. The LT segmentations shown in Figure \ref{fig_PNM_porosity_whole_core} match the results from \cite{Jackson2020} well, and give an estimate of the macro-porosity of each system. In each case the clay content in the core is $\approx$0.02 - 0.03 in porosity, in line with physical measures of Bentheimer sandstone \cite{Peksa2015}. The absolute segmentation threshold values are slightly different for the low resolution and super resolution images due to different feature representation and filtering in each image. Although filtering was optimised to give the highest SSIM, the noise distribution is inherently different between the different images. 

The LT and HT segmented subvolumes for the LR and SR whole core images are then passed to the pore-network modelling workflow as detailed in the previous section. Petrophysical properties are predicted for each subvolume, namely: porosity, permeability, relative permeability, and capillary pressure using the same fluid properties as before. This is performed in a highly distributed manner making use of parallel computing architecture - each subvolume is distinct and can run independently.  

With the known image and subvolume geometry, we create a 3D model of each core with the  subvolume petrophysical properties defined uniquely for each cell in the model. The model forms the base to conduct continuum scale multiphase simulations using the approach developed by the authors and previously presented in \cite{Jackson2018, Jackson2019, Jackson2020, Zahasky2020}. We note that each 3D cell (of cubic side length $\approx$ 1.5mm) in the model has unique, distinct, petrophysical properties deterministically predicted from the underlying image pore-network; there is no calibration or `history-matching' performed. 

With the 3D model defined, we numerically solve conservation of mass and momentum (two-phase Darcy's law) using the fully implicit, isothermal immiscible multiphase porous media flow simulator CMG IMEX\textsuperscript{TM}. Constant fluid properties and temperatures are used throughout the simulations. We use boundary conditions equivalent to those in the experiment - zero flux normal to the outer diameter of the core, a constant volume flux at the inlet face and a constant pressure at the outlet face. We vary the inlet flux to match the experimental fractional flows, and run the simulations for the same time periods until steady-state is achieved. We can then compare the resulting saturations in each cell to those in the experiment on a 1:1 basis, as well as core-averaged pressure drops in the form of absolute and relative permeability. 

\section{Results}

Here we present results from the multi-scale characterisation and modelling workflow. We first evaluate the performance of the deep learning algorithm in generating realistic super-resolution images from low-resolution images, using direct image comparisons and pore-network modelling. We follow this verification exercise with validation using whole core continuum simulations to compare directly to experiments across a variety of conditions.  

\subsection{Pore-scale results}
\label{sec_results_pore_scale}

\begin{figure*}
\subfloat[]{\includegraphics[width=0.95\textwidth]{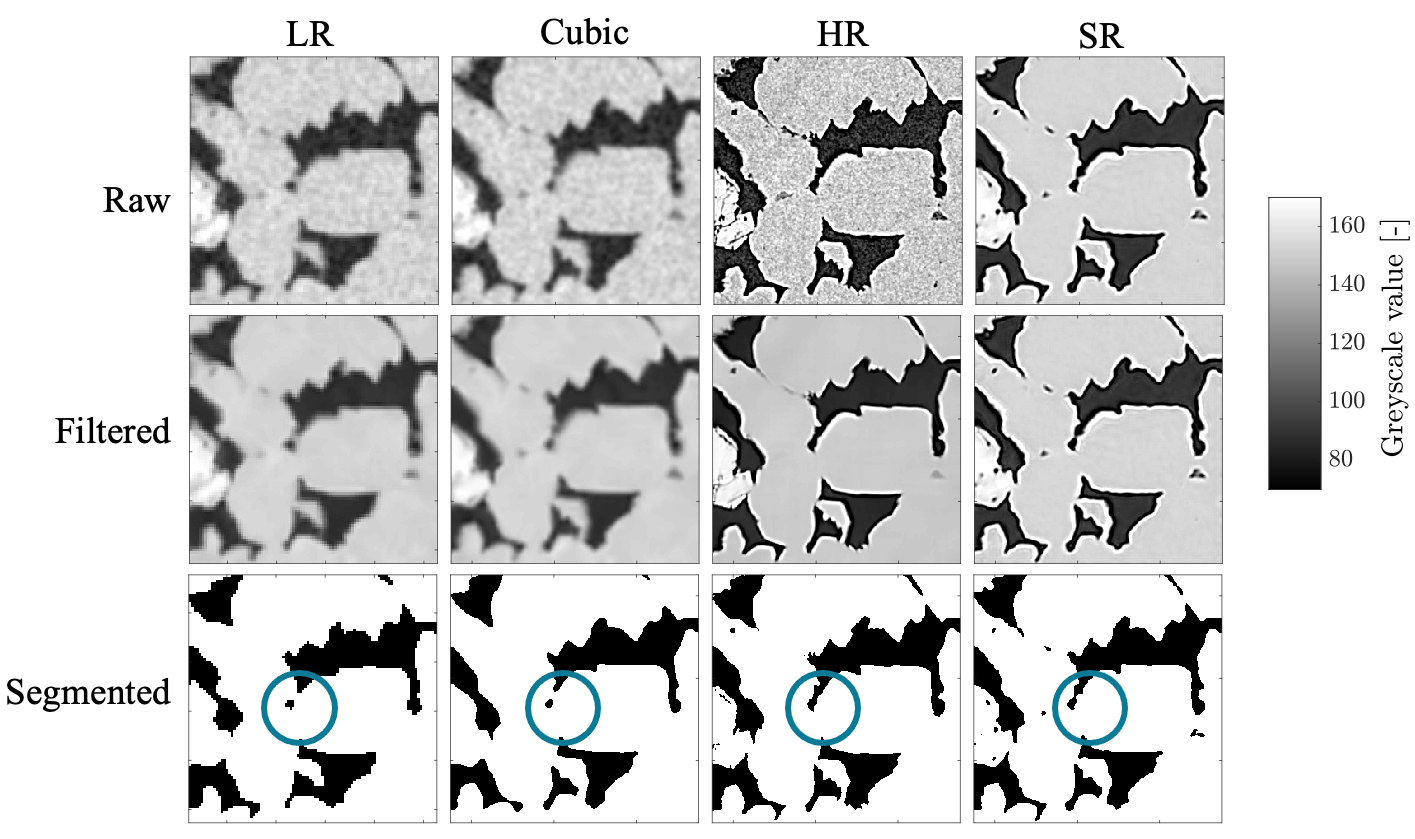}}

\subfloat[]{\includegraphics[width=0.95\textwidth]{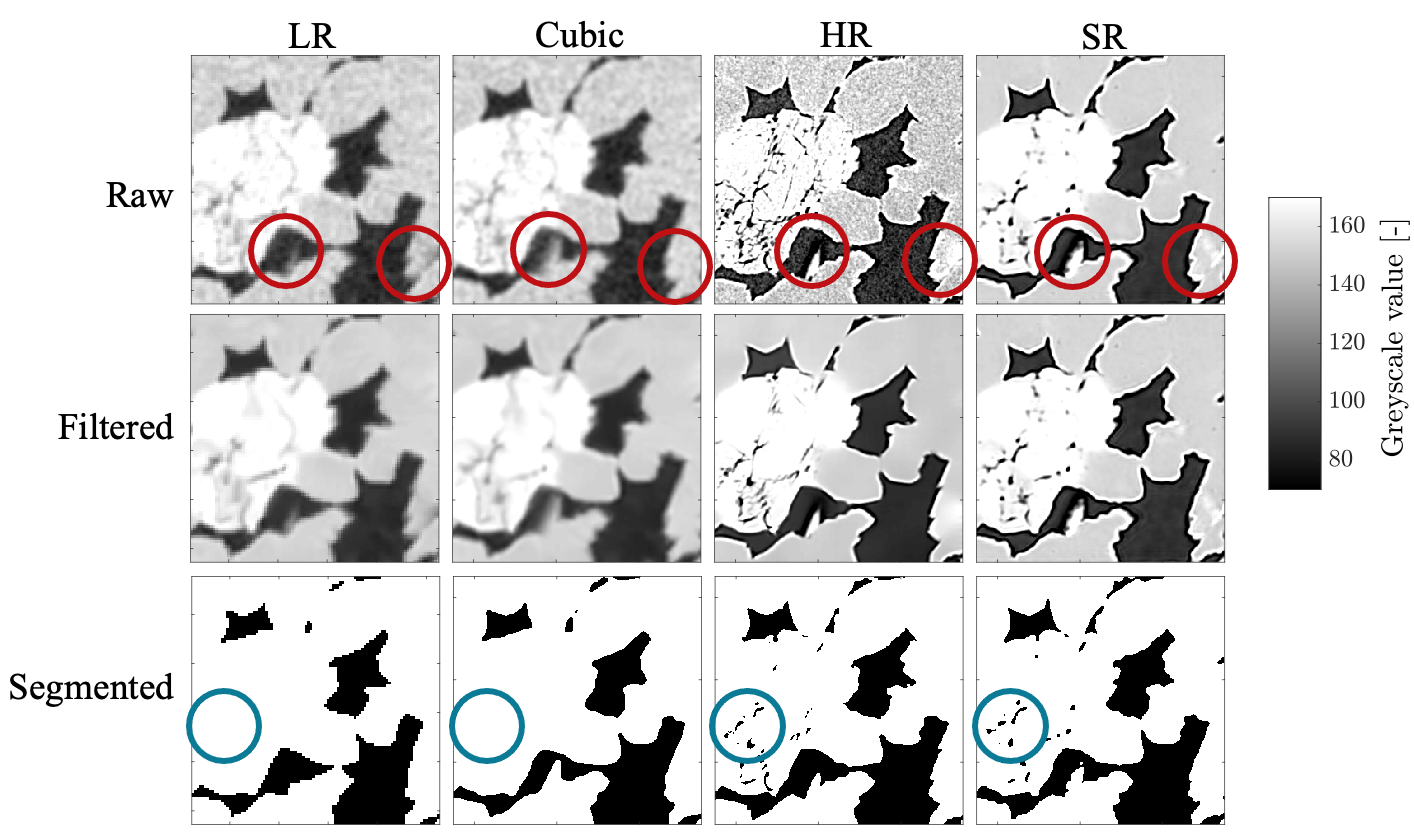}}
\caption{Comparisons of the LR, Cubic, HR and SR images. Region shown is a 600 x 600 $\mu$m 2D crop from the first slice, centred on pixel 175/225. (a) core 1 subvolume 2. (b) core 2 subvolume 2. In each sub figure, the top row are raw, normalised images, the middle row are filtered images and the bottom row are the segmented images using the base threshold of 117 grey scale value. The columns from left to right are LR, Cubic, HR and SR images, respectively. Blue circles highlight areas of interest showing differences between the images.The red circles show potential hematite mineral inclusions in core 2, which are also visible in many other regions (see the supporting information for further images).}
\label{fig_LR_HR_SR_comps}
\end{figure*}

We first discuss the pore-scale efficacy of the workflow. The first slice of the LR, Cubic, HR, and SR images for core 1 subvolume 2 and core 2 subvolume 2 are shown in Figure \ref{fig_LR_HR_SR_comps}. Only core 1 subvolume 1 was used in the training and testing of the deep learning algorithm; the images in Figure \ref{fig_LR_HR_SR_comps} are therefore completely unseen. Further image comparisons are shown in the Supporting Information Figures S1 - S3. We show the raw images as well as filtered images, which highlight the high filtering similarity that is achieved between the SR and HR images (e.g. Equation (\ref{eq_SSIM})). In Figure \ref{fig_LR_HR_SR_comps}, we see that the SR image, generated through the EDSR network, is able to reproduce many of the sharp, angular features apparent in the HR image, highlighted by the blue circle regions. The LR and Cubic images have much smoother feature representations, especially after filtering to the same level as the SR image. Upon segmentation, features are generally more rounded; the sharp grain-pore contacts typical of the Bentheimer sandstone are somewhat lost. 

\begin{figure*}[ht!]
\hspace*{-40pt}
\includegraphics[width=1.15\textwidth]{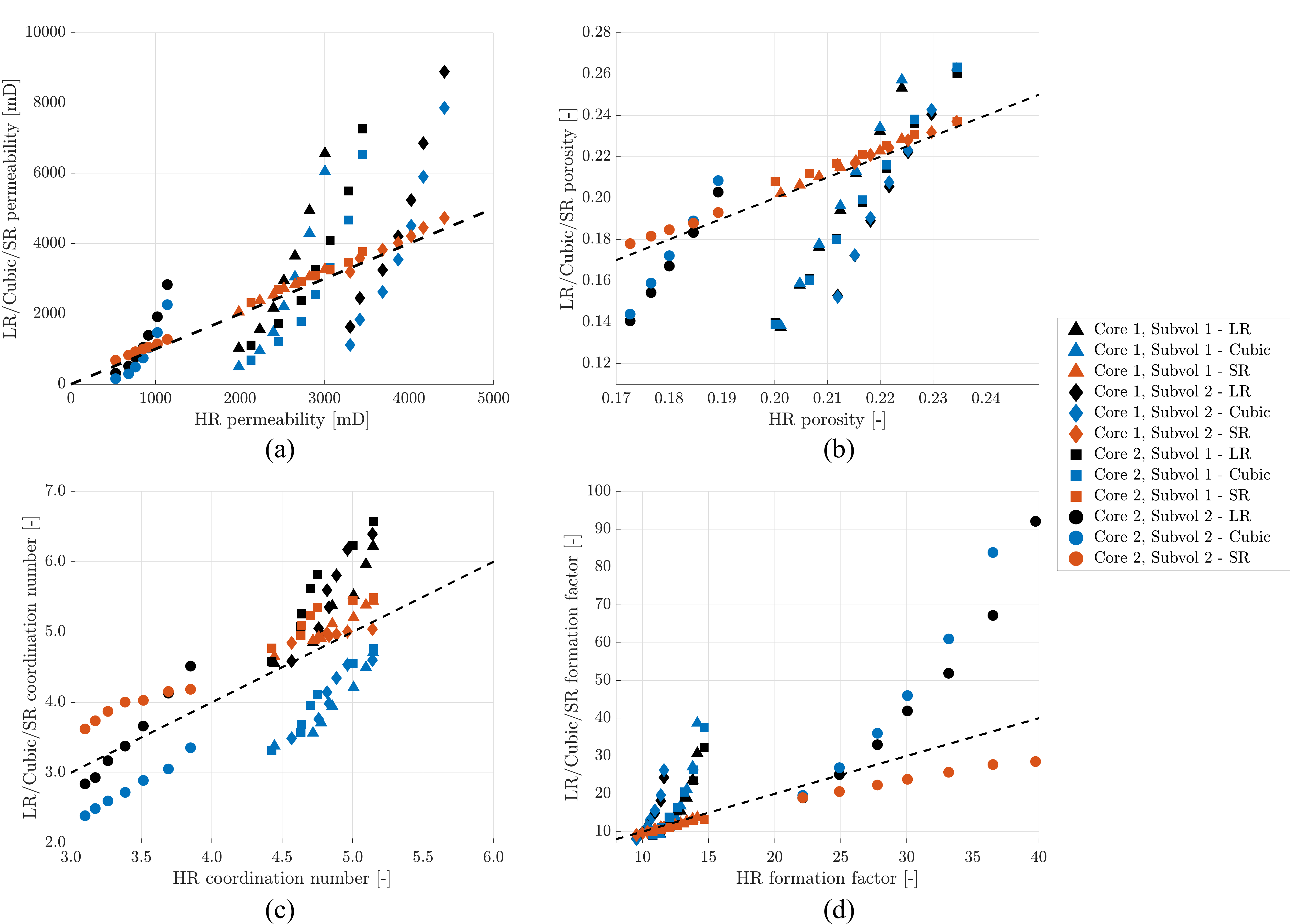}
\caption{Pore-network model simulations for the LR, Cubic, HR and SR images for all subvolumes. (a) Permeability. (b) Porosity. (c) Coordination number. (d) Formation factor. Dashed black 1:1 line shows the perfect correspondence of the LR/Cubic/SR with the HR data. Legend is shown on the RHS; symbol colours differentiate LR/Cubic/SR data, symbol types differentiate cores and subvolumes. Multiple points with the same colour and symbol are different segmentation realisations (7 in total).}
\label{fig_PNM_all}
\end{figure*}

As well as sharper grain-pore contact features, high frequency features are also significantly better represented in the SR image. Small pores and connecting throats are more resolved; this is apparent even when comparing to the Cubic image (at 2$\mu$m resolution), which is not able to capture the same detail. The EDSR has learnt a deeper mapping from the LR to HR image than a polynomial interpolation, using higher level information such as the image gradient, which improves small-feature representation. This is particularly important for the overall connectivity in the domain, and the subsequent pore-network extraction. Small disconnected regions (such as that shown in Figure \ref{fig_LR_HR_SR_comps}a, blue circle) could be wrongly attributed to a different pore, and may not be connected to the main flow path, leading to inaccuracies in flow simulation. Generally, the SR image appears more connected than the LR or Cubic images, in line with the HR image. 

We see that the contrast between features is also higher in the SR image compared to the LR/Cubic image. This is visible in the blue circle in Figure \ref{fig_LR_HR_SR_comps}b, whereby intra-granular porosity is more clearly visible, and retained upon segmentation in both the HR and SR images. Further analysis of the core 2 subvolume results in Figure \ref{fig_LR_HR_SR_comps}b (and in the SI) reveals the ability of the EDSR to learn specific mineralogy aspects of the images. Core 2 features bright patches of highly attenuating iron-oxide deposits in the form of hematite ($\approx 9 \times$ more attenuating than quartz \cite{Bam2020}). These also appear in the visible spectrum as a distinct red colour. These iron oxide deposits typically form around fault planes from iron-rich groundwater flow, and can form between quartz grains \cite{Dubelaar2015}. The SR image is able to reproduce these bright patches, even though they are barely visible in the relatively dark LR image (see Figure \ref{fig_LR_HR_SR_comps}b red circles). These patches can reduce the overall porosity/permeability of the sample when in-between grains and are key in the heterogeneity characterisation - see the simulation results for further explanation. 

The EDSR network was optimised for the micro-CT data here by reducing the number of residual blocks and filters in each layer; an optimum was manually found that produced low $L_1$ loss whilst not overfitting the data. Overfitting was clearly visible in early attempts when specific beam hardening and ring artefacts were learnt from the HR image. These are common X-ray micro-CT imaging artefacts, due to the non-linear attenuation of the core-holder/rock, and through dead pixels, respectively. The ring artefacts are visible in some HR images here (see the central region crop in Figure S1c,g,k,o), but have not been learnt by the EDSR. This is a key attribute of the EDSR implementation, and allows our SR images to represent the true features of the HR images whilst removing common, and often unavoidable imaging artefacts. 

A further feature that the EDSR network has learnt is the grain-edge artefacts apparent in the HR image. These bright regions are due to small-angle X-Ray refraction (often interpreted as `phase-contrast' artefacts), and occur increasingly at high resolution and high energy with sharply contrasting features \cite{Lange2012}. They are not necessarily a negative effect in dry images, since they increase contrast for segmentation, however they can be negative if multiple (fluid) phases are present in the sample, especially if phase contact geometry is required. These features are not learnt through overfitting, since they are ubiquitous in the HR image here; they could be removed somewhat with phase contrast removal techniques. 

It is worth noting that in Figure \ref{fig_LR_HR_SR_comps} we only show one segmentation result, at the base threshold value (grain-pore minima value 117), to give a representative view of the pore-space. Although other threshold values, which are not necessarily equal to each other, may result in closer visual matches between comparison images, the base threshold shown highlights the general trends between the LR, HR, and SR images. To more quantitatively compare the images, we now discuss the results from the pore-network extractions and flow simulations.  

Figure \ref{fig_PNM_all} shows petrophysical property predictions from the pore-network model extractions and flow simulations comparing the LR/Cubic/SR to the HR result. In each case, 7 segmentation results are shown with the middle symbol representing the base case in each case. As the segmentation threshold is increased, the grain space is eroded and the porosity/permeability is increased. Generally, it is clearly visible that the SR results closely match the HR results across all segmentation thresholds, closely following the unity line. This is not true for the LR and Cubic results, which generally match the HR result at only one segmentation threshold, which is not equal to that of the HR image. The crossover point for the LR/Cubic results with the unity lines changes for each petrophysical property;  therefore, it is not clear which segmentation threshold for the LR images is optimal compared to the HR image. This highlights the potential pitfalls with choosing a segmentation for a lower resolution image based on a single criteria \cite{Garfi2019}, which does not necessarily hold for other petrophysical predictions. The SR image on the other hand displays consistent behaviour with the HR image across the range of segmentations. This means that given an external measurement to calibrate against, the chosen segmentation threshold should lead to consistent petrophysical property prediction, in line with the HR image. Furthermore, the variance of the SR petrophysical properties with segmentation choice is less than the LR data, indicating it is less sensitive to the segmentation choice itself. 

Out of the 4 subvolumes shown in Figure \ref{fig_PNM_all}, core 2 subvolume 2 has the least consistent SR result compared to the HR image. This is likely due to the quantitivative difference in the pore-structure between that subvolume and the EDSR training subvolume (core 1, subvolume 1). The subvolume 2 location has high levels of the bright hematite mineral, and a tight grain structure with a  permeability around one order of magnitude less than core 1 subvolume 1. However, the petrophysical predictions are still acceptable, and generally within $\pm$ 5-10\% of the HR image results. In this work, we consciously chose to train the EDSR network on only one subvolume in one core, to highlight the small training dataset that is required. The training size was chosen based on a typical REV of Bentheimer, and chosen randomly from the datasets. If the training batch were extended to include multiple regions, then conformance could likely be achieved over data with more variance. As in all deep learning methodologies, the performance of the network is strongly tied to the quality/quantity of the training dataset, and its representativeness of the test data \cite{da2021deep}.

\begin{figure}
\subfloat[]{\includegraphics[width=0.42\textwidth]{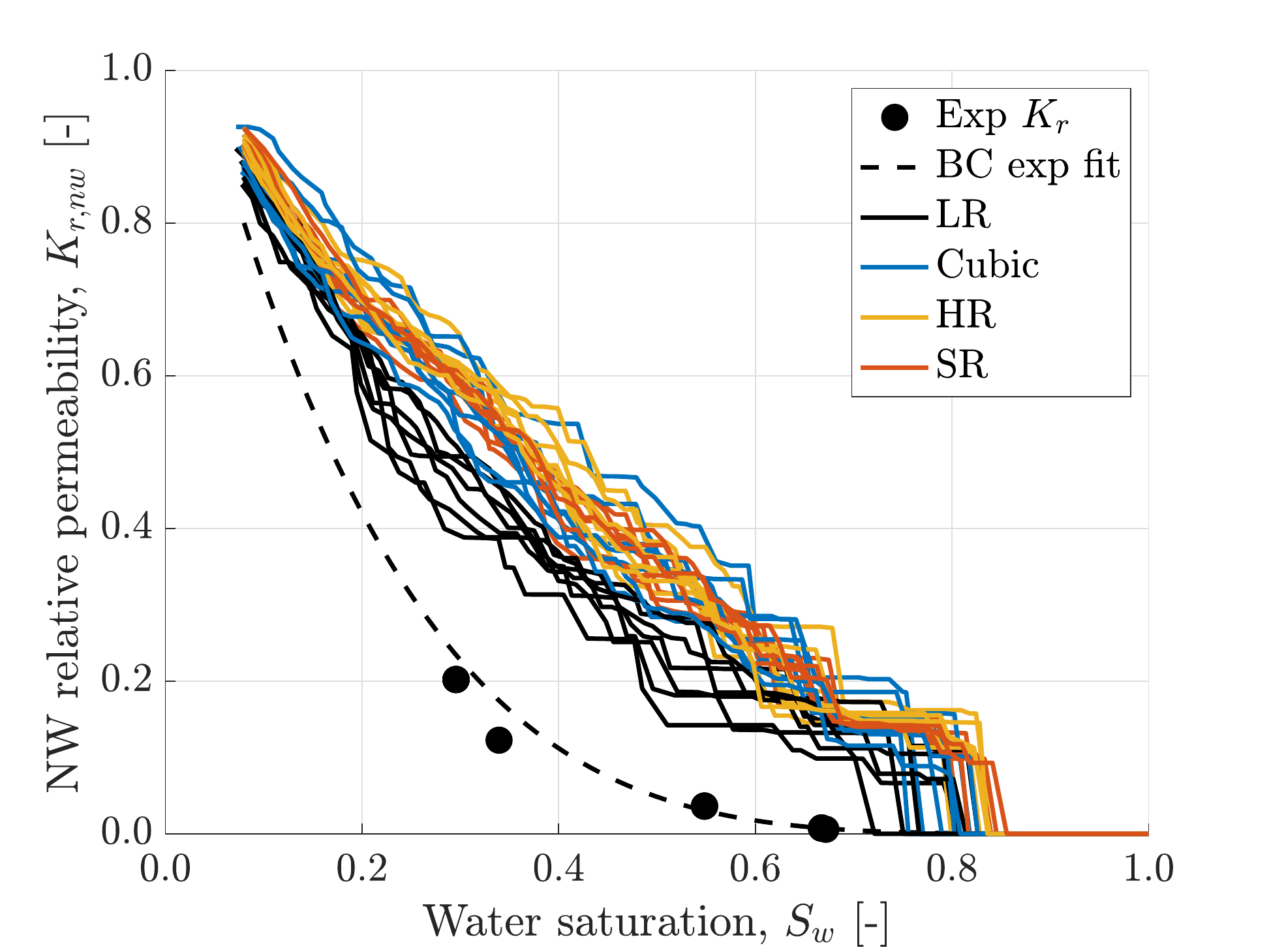}}

\subfloat[]{\includegraphics[width=0.42\textwidth]{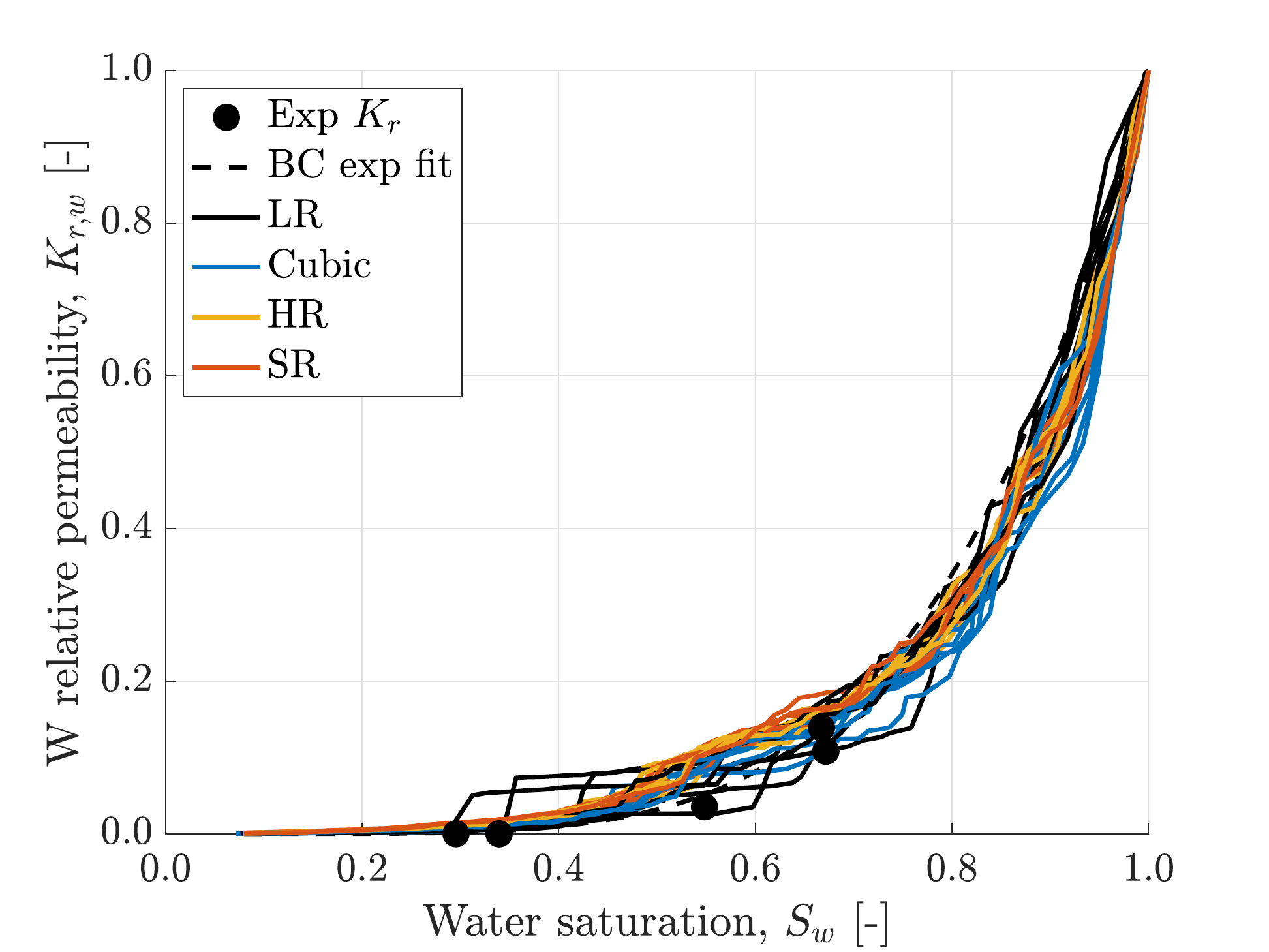}}

\subfloat[]{\includegraphics[width=0.42\textwidth]{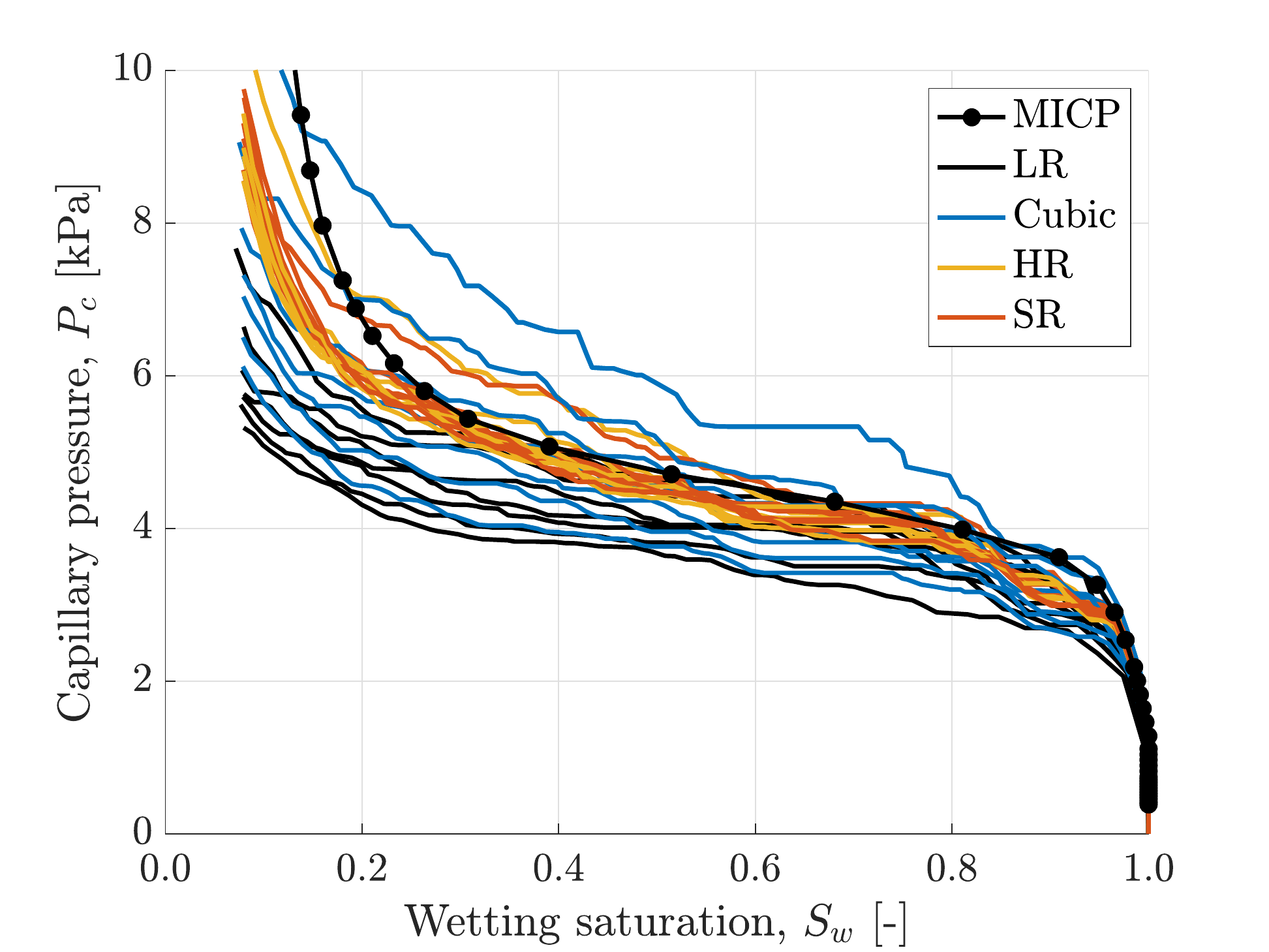}}
\caption{Pore-network model simulation results for the LR, Cubic, HR and SR image networks of core 2 subvolume 1. Each line represent a flow simulation on one of seven different segmentation thresholds from -15\% to +15\% around the base threshold of 117 greyscale value. (a) Non-wetting relative permeability. (b) Wetting relative permeability. (c) Capillary pressure. The MICP data is from a sister core. The exp Kr data is from the core averaged multiphase flow experiments. There is a Brooks-Corey (BC) function fit to the experimental relative permeability data also shown from \cite{Jackson2020}.}
\label{fig_PNM_kr}
\end{figure}

Multiphase flow predictions from the pore-network models are shown in Figure \ref{fig_PNM_kr} for core 2 subvolume 1 (other subvolume results are shown in the SI Figure S4-6). The HR and SR results both show less variation across the different segmentation thresholds than the LR/Cubic result, and closely match each other generally. In particular, the capillary pressure predictions for the SR and HR images closely match the externally derived mercury intrusion capillary pressure (MICP) data, when scaled to the correct IFT fluid pair \cite{pini:2012pi}. Higher capillary pressures at $S_w<0.2$ are better represented in the SR and HR images, due to the resolution of small pores and throats in the system. In the LR/Cubic networks, these small pores are poorly resolved, and hence the maximum capillary pressure that can be simulated is limited. We also see that the gradient of capillary pressure ($|dP_c/dS_w|$) in the low capillary pressure region ($0.2 < S_w < 0.8$) is smaller for the LR/Cubic networks compared to the SR and HR networks and MICP data. A low $|dP_c/dS_w|$ across a range of $P_c$ values is indicative of a narrow pore-size distribution, where many pores are filled for small changes in $P_c$. The LR/Cubic networks have a narrower range of pore-sizes compared to the HR and SR data, as well as the real rock, misrepresenting the filling process and capillary pressure in the system. Further, we see that the SR and HR networks have a better representation of the threshold capillary pressure at the first inflection point of $dP_c/dS_w$, where percolation across the system starts. The more connected domains generally in these systems allow better representation of the initial filling stages. 

The SR and HR relative permeability predictions in Figure \ref{fig_PNM_kr}a, b also show less variance compared to the LR/Cubic predictions. The wetting relative permeability is generally well predicted by all networks, with little variation across the segmentations. Since the wetting phase is connected across the domain for all saturations, the relative permeability is quite insensitive to pore-structure variations created by the different network morphologies. However the wider, more connected pore-size distribution of the SR and HR networks does result in smoother transitions between filling states, with less discrete `jumps' than the LR/Cubic network. The non-wetting permeability shows larger variations across the networks, especially for the LR/Cubic images at the percolation threshold. Variations in the network morphology have stronger impacts on the non-wetting percolation, especially given the pore-space heterogeneity. We note that all networks over-predict the experimental core averaged values here. This is largely due to finite-size impacts, as the smaller subvolume networks are more readily percolated than the whole core system. The REV size used in this work is the same as \cite{Jackson2020} and chosen based on porosity and capillary pressure. It is possible that relative permeability REVs are larger. Further from this, capillary-end effects are present in the whole core (and not in the PNM), which generally reduce relative permeabilities and are present in the effective experimental $K_r$ values; this is discussed in more detail in the continuum modelling section \ref{sec_results_continuum_scale}. These results further highlight potential user bias when evaluating results compared to a single criteria; care must be taken to consider scale and representativeness. 

Results from the other subvolumes in the SI highlight the pore-scale heterogeneity across the rock samples, and the general ability of the SR networks to capture this in line with the HR networks. The varying capillary pressure compared to the base MICP data is a direct consequence of the changing pore-structure and therefore capillary pressure heterogeneity. 

In this section we have rigorously validated the ability of the EDSR network to produce physically realistic SR images of the pore-space from LR images. We have compared common image similarity metrics (SSIM), visual texture, and flow simulations to the corresponding LR and HR images. Distinct from previous works \cite{Wang2019, Wang2019b, Wang2020, Janssens2020, Niu2020, Niu2021}, we have developed the EDSR network in 3D and demonstrated the pore-scale validation across multiple segmentation realizations from multiple subvolumes in different samples, gaining a thorough understanding of the uncertainty. We also perform multiphase flow validations, which are crucial for subsurface applications, but are often lacking in previous, less application driven studies. Further from previous work, we also validate the results with true low and high-resolution images obtained from optical magnification of the samples, rather than artificial up/downscaling of a single image with numerical procedures. This is common in earlier verifications of the approaches, but does not provide a true representation of noise transfer with scale in realistic applications \cite{Wang2019b}. 

\subsection{Continuum-scale results}
\label{sec_results_continuum_scale}

The verification in the previous section was performed primarily with numerical simulation; we now provide direct experimental validation of the EDSR generated images and modelling approach. As described in section \ref{sec_methods_cont_model}, we use the trained EDSR network to generate high-resolution subvolumes across the full core samples from the low-resolution images. These subvolumes are segmented, and fed into the PNM to generate petrophysical properties, and ultimately a 3D full core continuum model. These models permit continuum flow to be simulated by solving conversation of mass and momentum, which we compare to the experimental results from \cite{Jackson2020}. We note the experimental results are from exactly the same rock samples, and hence can be compared to the model results here in a 1:1 approach. 

\begin{table}[ht]
\caption{Single phase flow simulation results compared to experiments. LT and HT refer to low threshold and high threshold generated models, respectively.}
\label{tab_single_phase}
\begin{tabular}{|l|c|c|c|c|}
\hline
\multicolumn{1}{|c|}{\multirow{2}{*}{\textbf{Property}}} & \multicolumn{2}{c|}{\textbf{Core 1}} & \multicolumn{2}{c|}{\textbf{Core 2}} \\ \cline{2-5} 
\multicolumn{1}{|c|}{}                                   & \textbf{LR}       & \textbf{SR}      & \textbf{LR}       & \textbf{SR}      \\ \hline
$K_{abs}$ experiment {[}D{]}                             & \multicolumn{2}{c|}{1.636}           & \multicolumn{2}{c|}{0.681}           \\ \hline
$K_{abs}$ LT {[}D{]}                          & 2.494             & 1.503            & 2.350             & 1.260            \\ \hline
$K_{abs}$ LT error {[}\%{]}                   & 52.5              & -8.1             & 245.1             & 85.0             \\ \hline
\begin{tabular}[c]{@{}l@{}}LT Dykstra-Parson \\coefficient, $V_k$ [-] \end{tabular}     & 0.174             & 0.280     & 0.213             & 0.285            \\ \hline
$K_{abs}$ HT {[}D{]}                         & 3.477             & 2.413            & 3.327             & 2.201            \\ \hline
$K_{abs}$ HT error {[}\%{]}                  & 112.6             & 47.5             & 388.5             & 223.2            \\ \hline

\begin{tabular}[c]{@{}l@{}}HT Dykstra-Parson \\coefficient, $V_k$ [-] \end{tabular}     & 0.152             & 0.202 & 0.202             & 0.221            \\ \hline

\end{tabular}
\end{table}

\begin{figure*}
\subfloat[]{\includegraphics[width=0.45\textwidth]{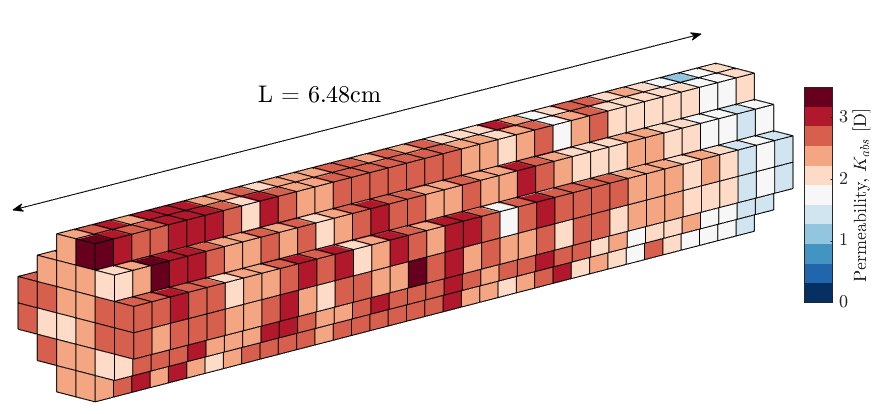}}\quad
\subfloat[]{\includegraphics[width=0.45\textwidth]{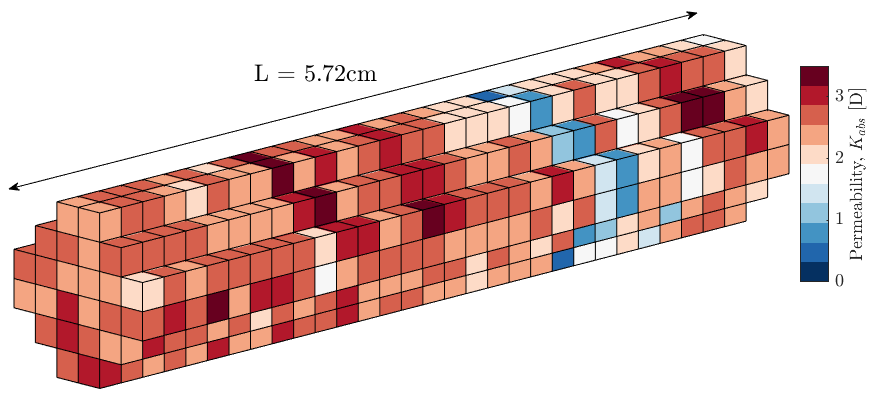}}

\subfloat[]{\includegraphics[width=0.45\textwidth]{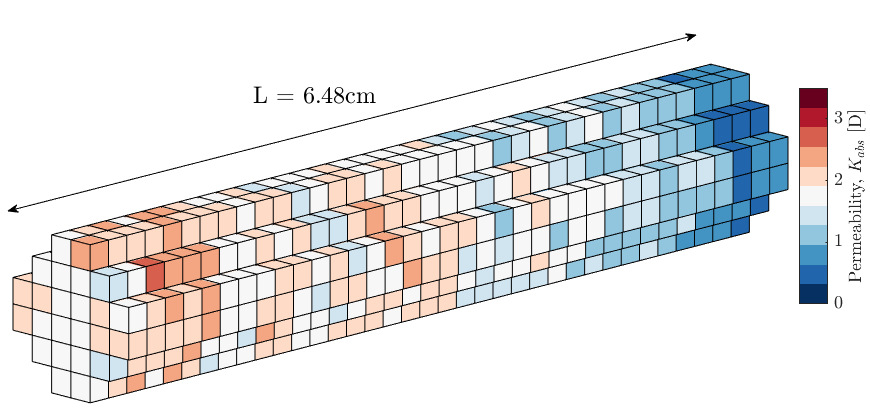}}\quad
\subfloat[]{\includegraphics[width=0.45\textwidth]{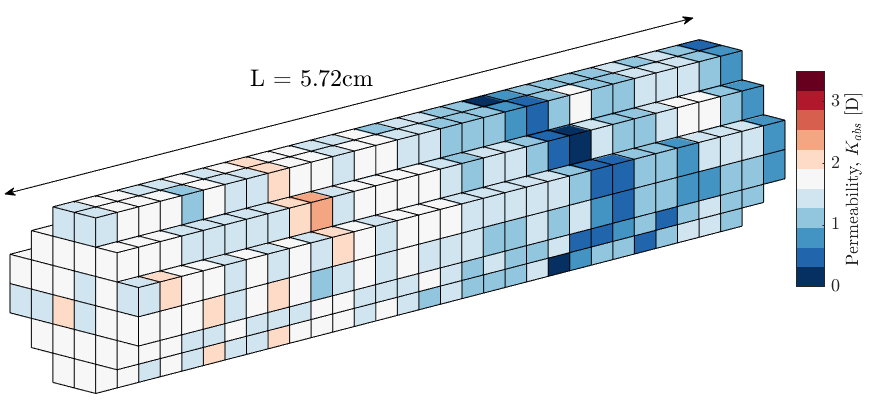}}

\subfloat[]{\includegraphics[width=0.45\textwidth]{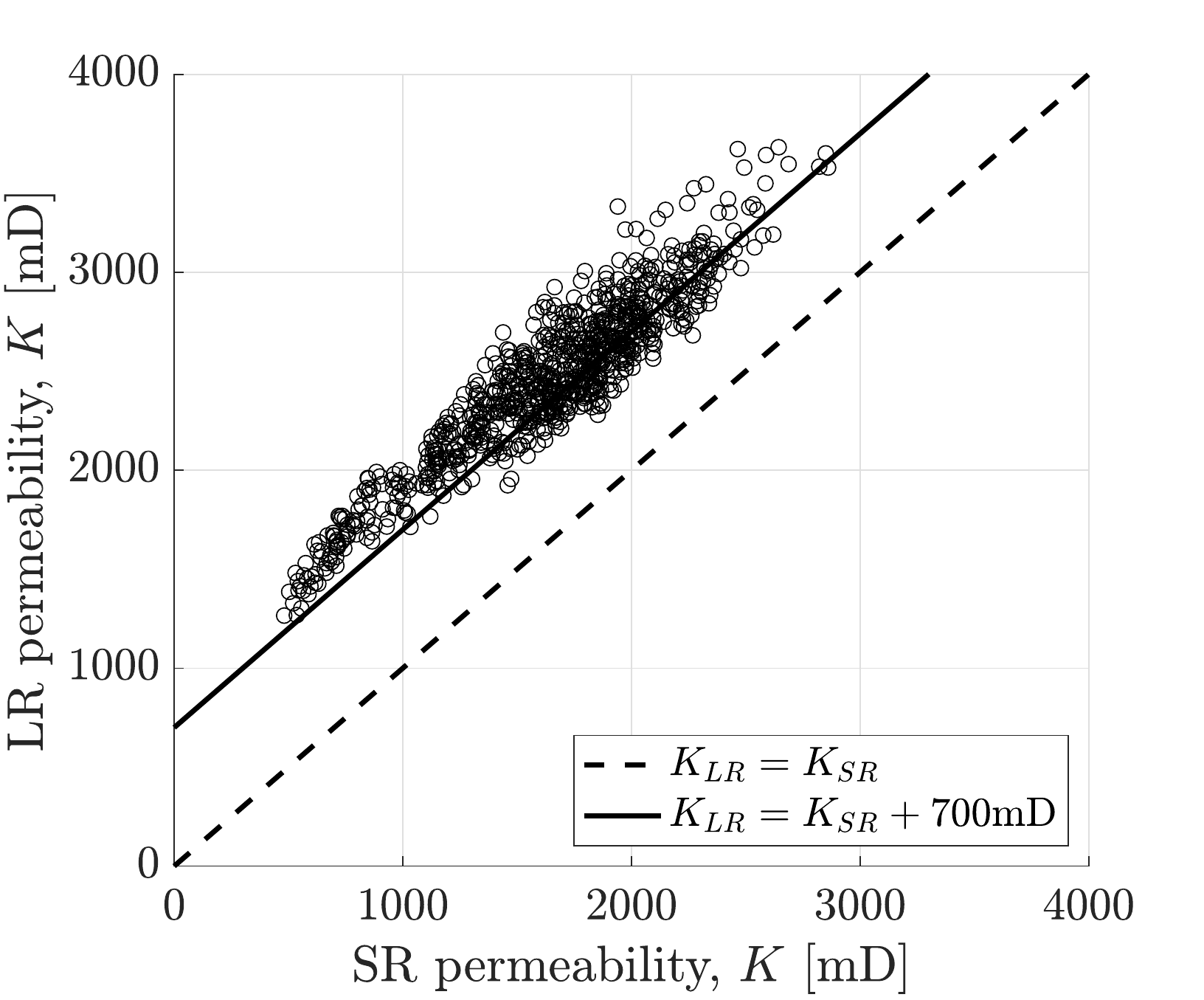}}\quad
\subfloat[]{\includegraphics[width=0.45\textwidth]{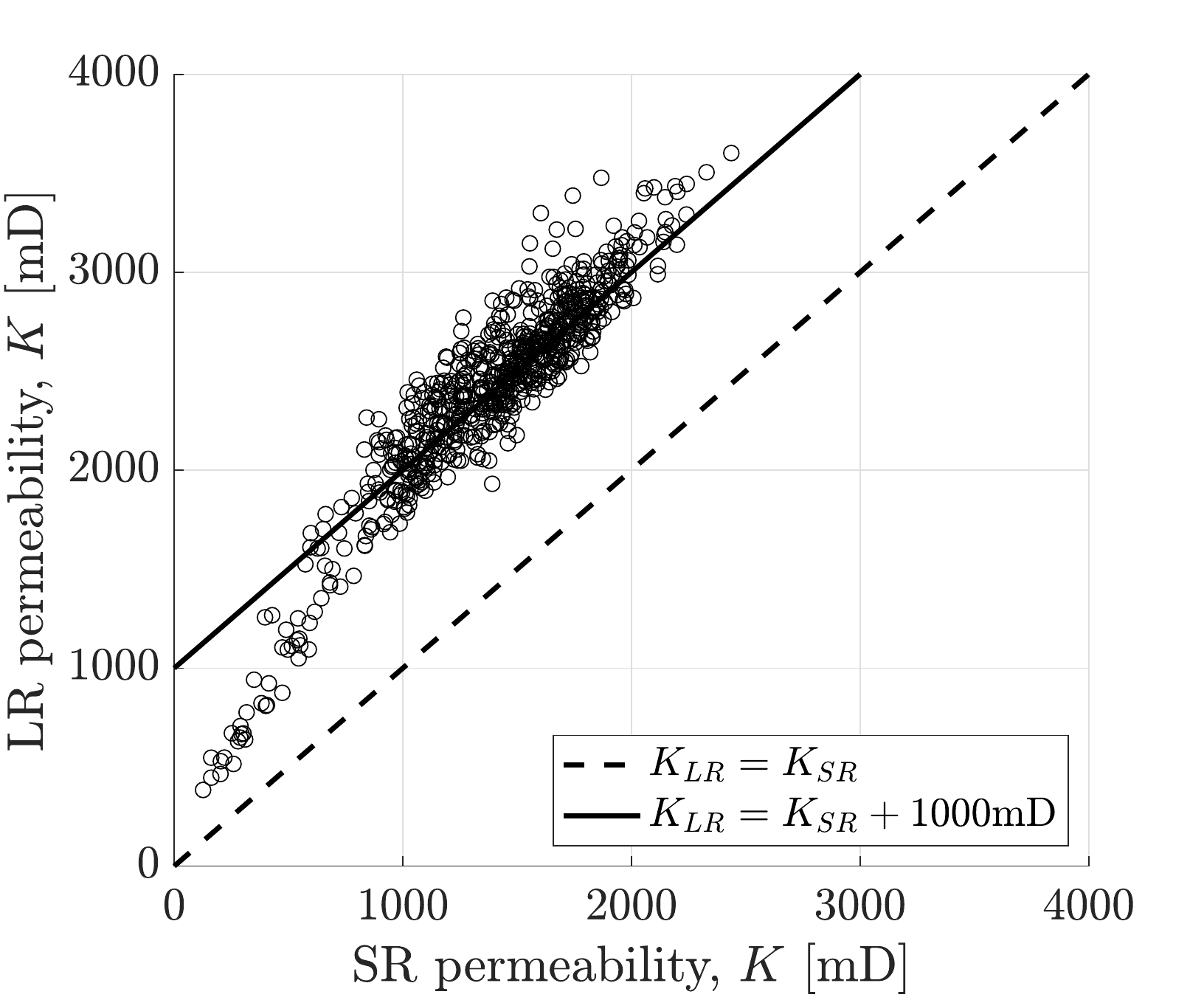}}

\caption{Continuum scale permeability results from the PNM using the LR LT and SR LT images.  (a, c) Voxelised permeability core 1, LR and SR models, respectively. (b, d) Voxelised permeability core 2 LR and SR models, respectively. HT images are found in SI Figure S7. (e, f) SR vs. LR voxel permeabilities for core 1 and core 2 respectively. }
\label{fig_whole_core_kr}
\end{figure*}

First, we compare core average absolute permeabilities. These are calculated in the experiments and simulation using the average pressure drop across the samples in an analogous manner to equation (\ref{eqn_perms}). Tabulated results are shown in Table \ref{tab_single_phase} for all model realizations, with 3D permeability maps for the LR and SR LT models displayed in Figure \ref{fig_whole_core_kr}a-d (further images are shown in SI Figure S7). We also show direct 1:1 comparisons of the voxel permeability predictions for the LR and SR LT images in Figure \ref{fig_whole_core_kr}e,f. For each core, we see that the SR model provides a more accurate prediction of the absolute permeability for both the LT and HT realizations when compared to the experimental average values. 

We show the Dykstra-Parson coefficient, $V_k$, in Table \ref{tab_single_phase}, which is the ratio of standard deviation in absolute permeability to the mean. We see that the SR results generally have higher $V_k$, indicating that they have captured more of the heterogeneity of the core. The HT segmentations generally result in lower $V_k$, as the pore-space has been more strongly eroded, with higher connectivity, resulting in  less permeability variation across the REVs. 

In Table \ref{tab_single_phase}, we see the HT realizations generally overestimate the permeability for both the LR and SR cases, since the pore-space has been eroded and enlarged. This increases local conductivity between pores, and the overall core averaged conductivity/permeability. At the lower threshold, the SR based model has an accurate prediction of the permeability for core 1 (8.1\% error); this threshold was chosen to approximate the macroscopic porosity in the sample that contributes to flow. When combined with the more accurate SR pore-space representation, this threshold gives a good estimate of the connected flow path and hence the permeability. However, for core 2, the SR LT model overpredicts the permeability (85.0\% error), although it is more accurate than the other models. For core 2, the 3D permeability map in Figure \ref{fig_whole_core_kr}d shows a very low permeability band cutting oblique to the flow direction, with permeability $O$(100)mD. This band is the primary control on the core average permeability (through the harmonic mean), with the rest of the core otherwise quite similar to the more homogeneous core 1 (i.e. $\approx$1.5D). Simulations show that reducing the permeability of cells within the band to 10\% of their original values (now ranging from 10mD - 50mD) reduces the average permeability to 793.2mD, in line with the experimental value (16.5\% error). These low values are likely linked to the hematite inclusions in the band, which reduce inter-granular porosity and permeability. It is likely that even the HR image of this region (the core 2 subvolume 2 data), at 2$\mu$m resolution, is still not able to fully resolve some small throats that are controlling the resistance to flow. With this, the SR image is also not able to capture the same features, and may need a higher resolution training dataset for capturing the very low permeability here. In previous works, models of resolution 2-5 $\mu$m have been used to accurately capture permeabilities of $>$100mD \cite{Mostaghimi2012, Botha2016}. However, for smaller permeabilities of $O$(10)mD, \cite{Callow2020} used resolutions $<1\mu$m, which may be required for the heterogeneous regions here. This highlights the care that must be taken when using real heterogeneous media that could have regions of very reduced permeability in a system of overall quite high average permeability.

When comparing the LR and SR results directly in Figure \ref{fig_whole_core_kr}e, f, we see that the LR results are highly correlated with the SR results at high permeability, but offset by a constant ($\approx$+700mD for core 1, $\approx$+1000mD for core 2). At permeabilities $<1$ Darcy, the results are less correlated; this suggests that the connectivity in the micro-structure is fundamentally different in the segmentations of the tight regions, and not as accurately captured by the LR image. At higher permeability the constant offset suggests that although porosity between the models are very well matched, the LR model is consistently not resolving some small features below a threshold. To achieve the same porosity, some of the larger regions are essentially eroded, resulting in larger conductivities. The permeability difference between LR and SR is near constant since these features are consistently missing in each LR image that represents an REV for permeability.  

We note that further reductions to the segmentation threshold value do not significantly lower the permeability predictions, which can actually inflect, and start increasing permeability past a given low threshold \cite{Janssens2020}. This is because lowering the threshold contracts the pores and throats, eventually entirely removing some small throats (which were greatly increasing resistance to flow). This results in flow through larger throats with increased permeability. This also helps to explain why the SR models generally have lower permeability predictions than the LR, since smaller throats are actually resolved, and can contribute to flow resistance in the network. 

\begin{figure*}
\subfloat[]{\includegraphics[width=0.45\textwidth]{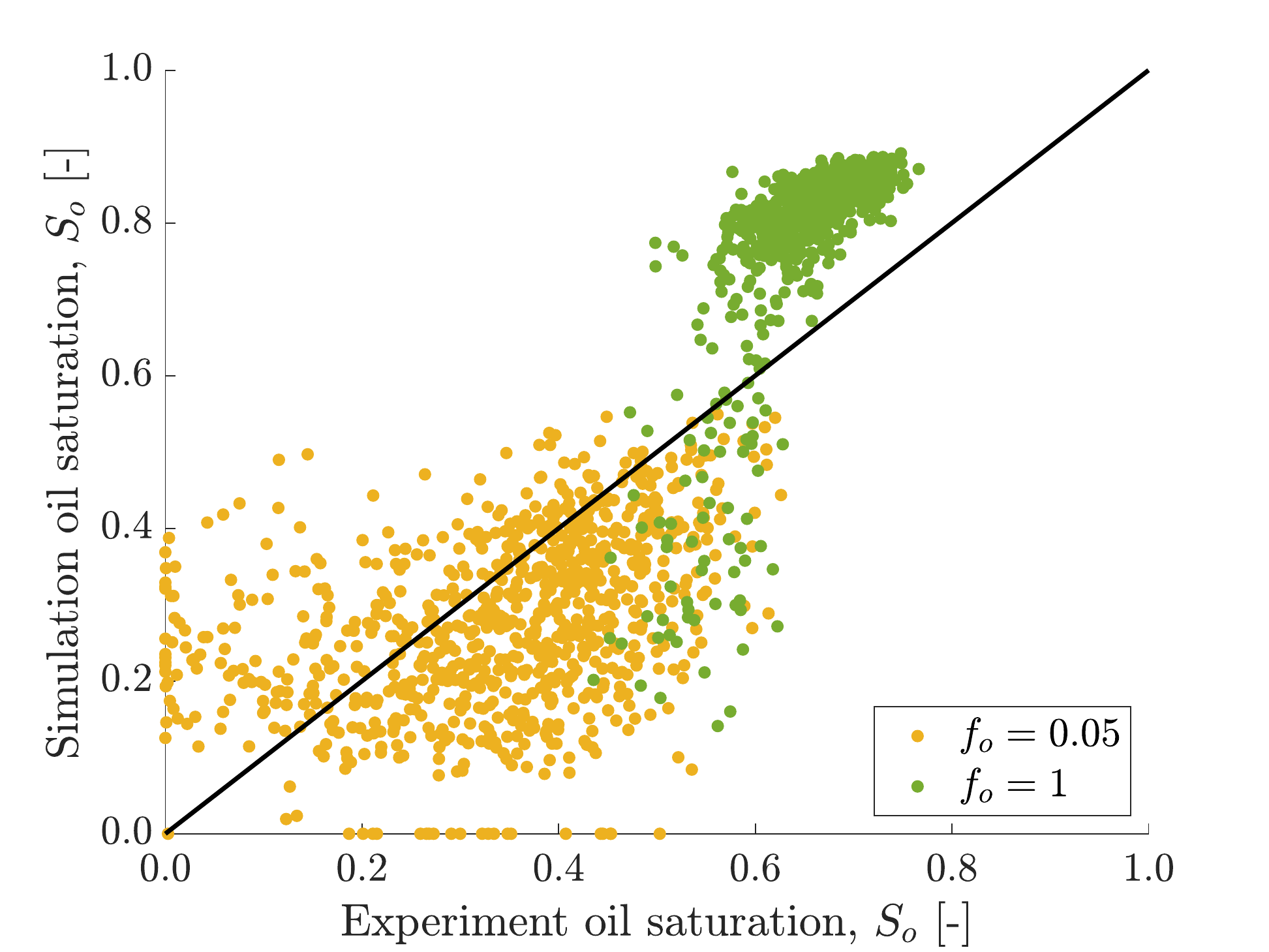}}
\subfloat[]{\includegraphics[width=0.45\textwidth]{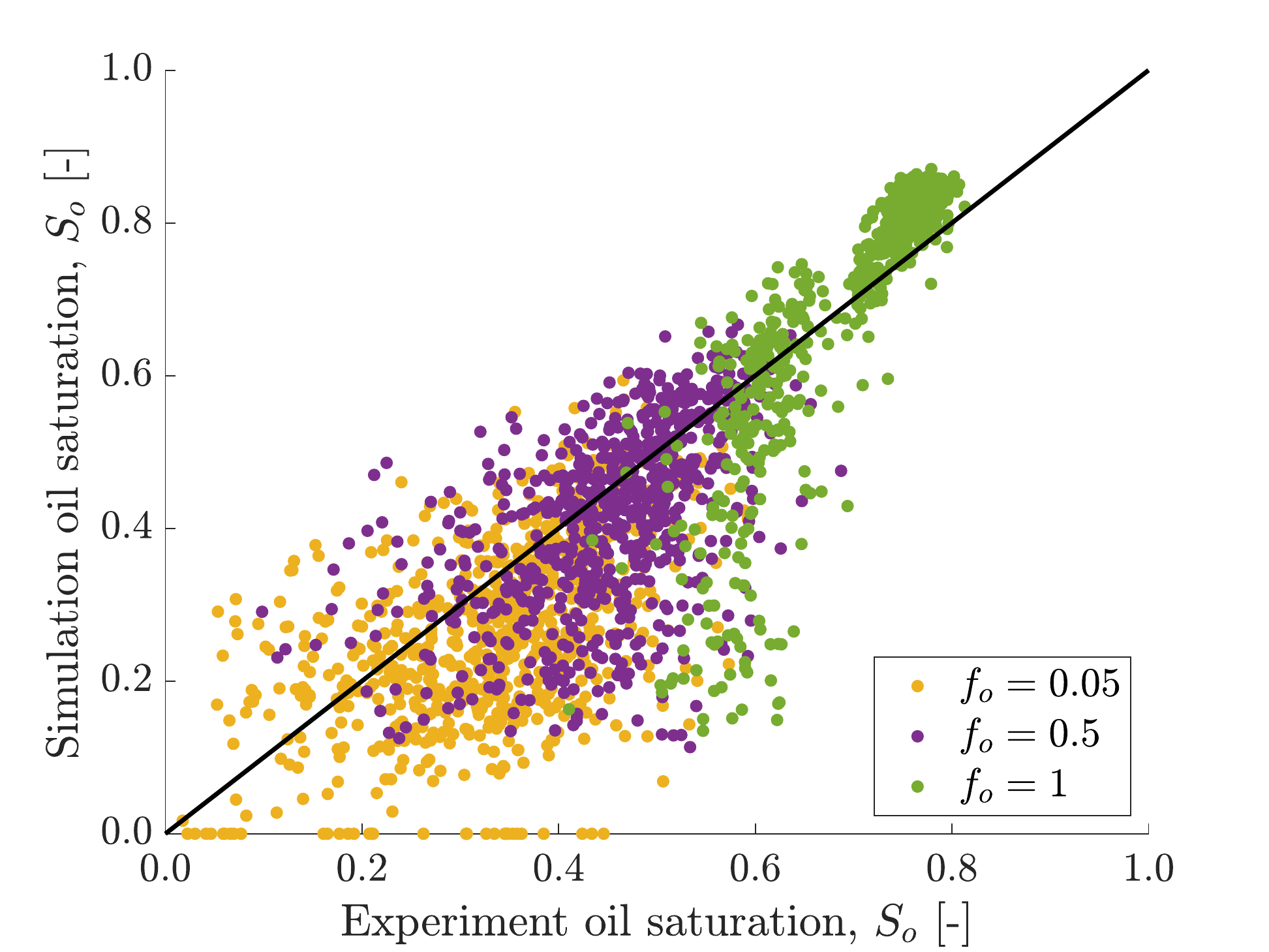}}

\subfloat[]{\includegraphics[width=0.45\textwidth]{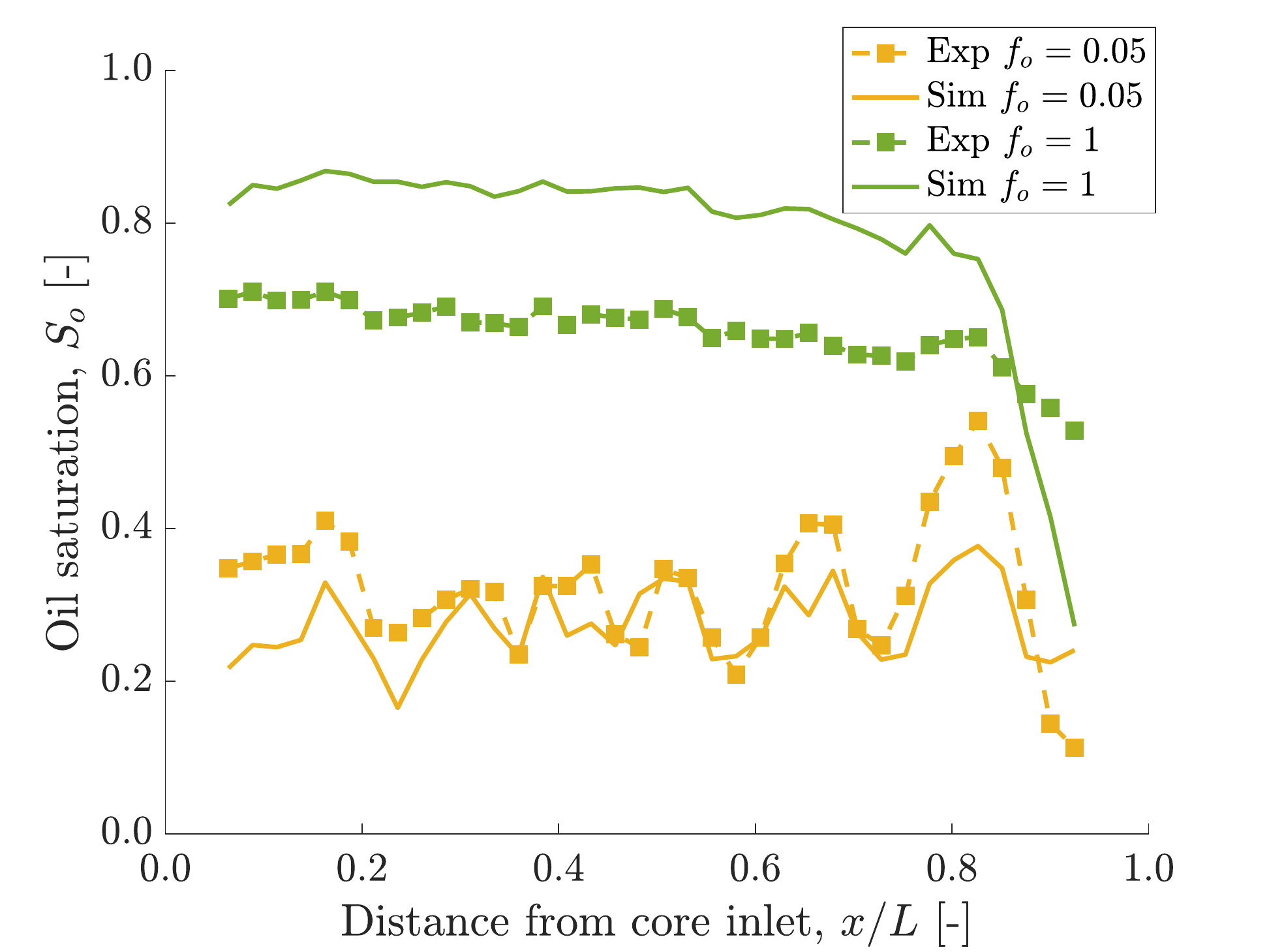}}
\subfloat[]{\includegraphics[width=0.45\textwidth]{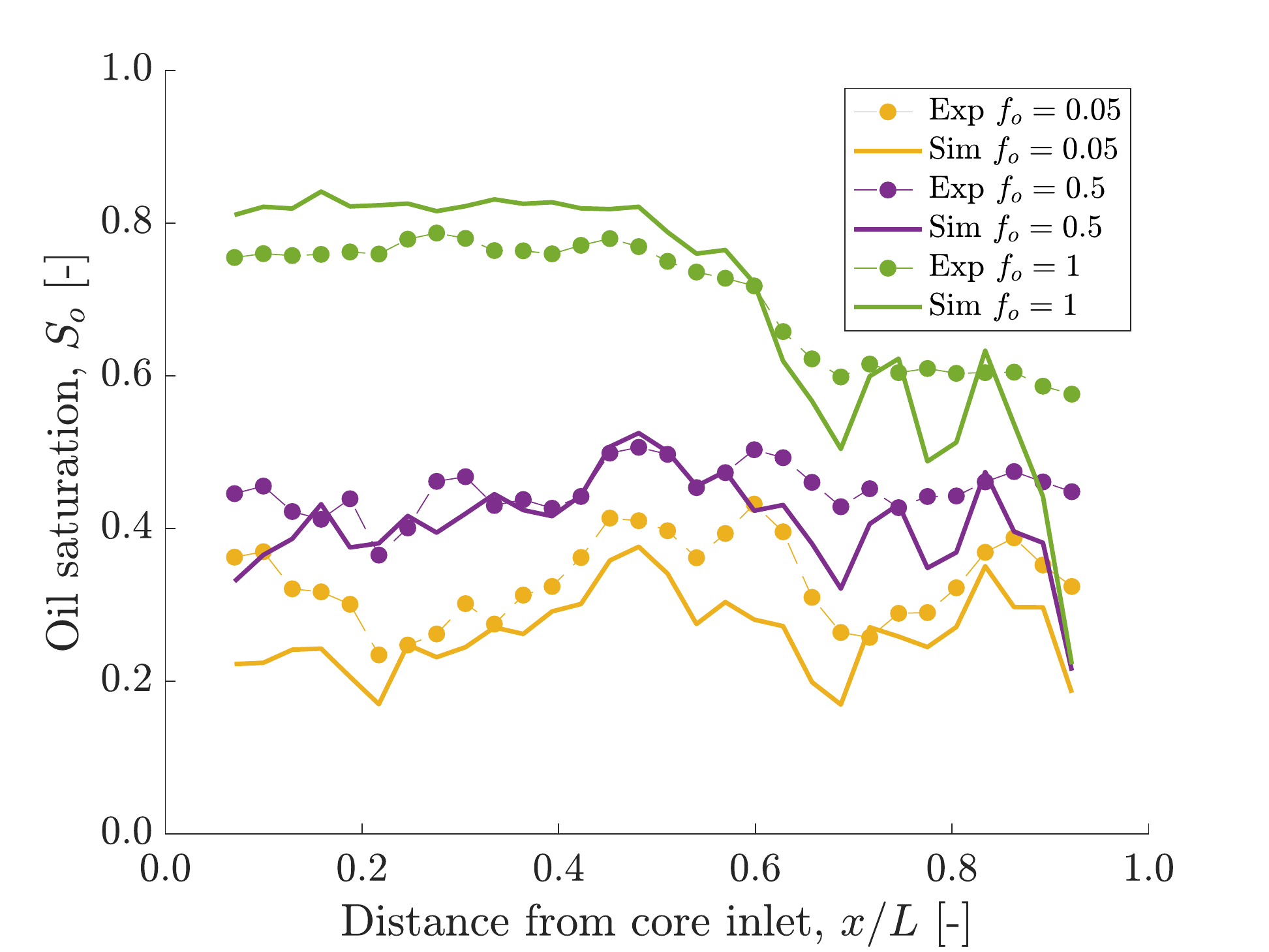}}

\subfloat[]{\includegraphics[width=0.45\textwidth]{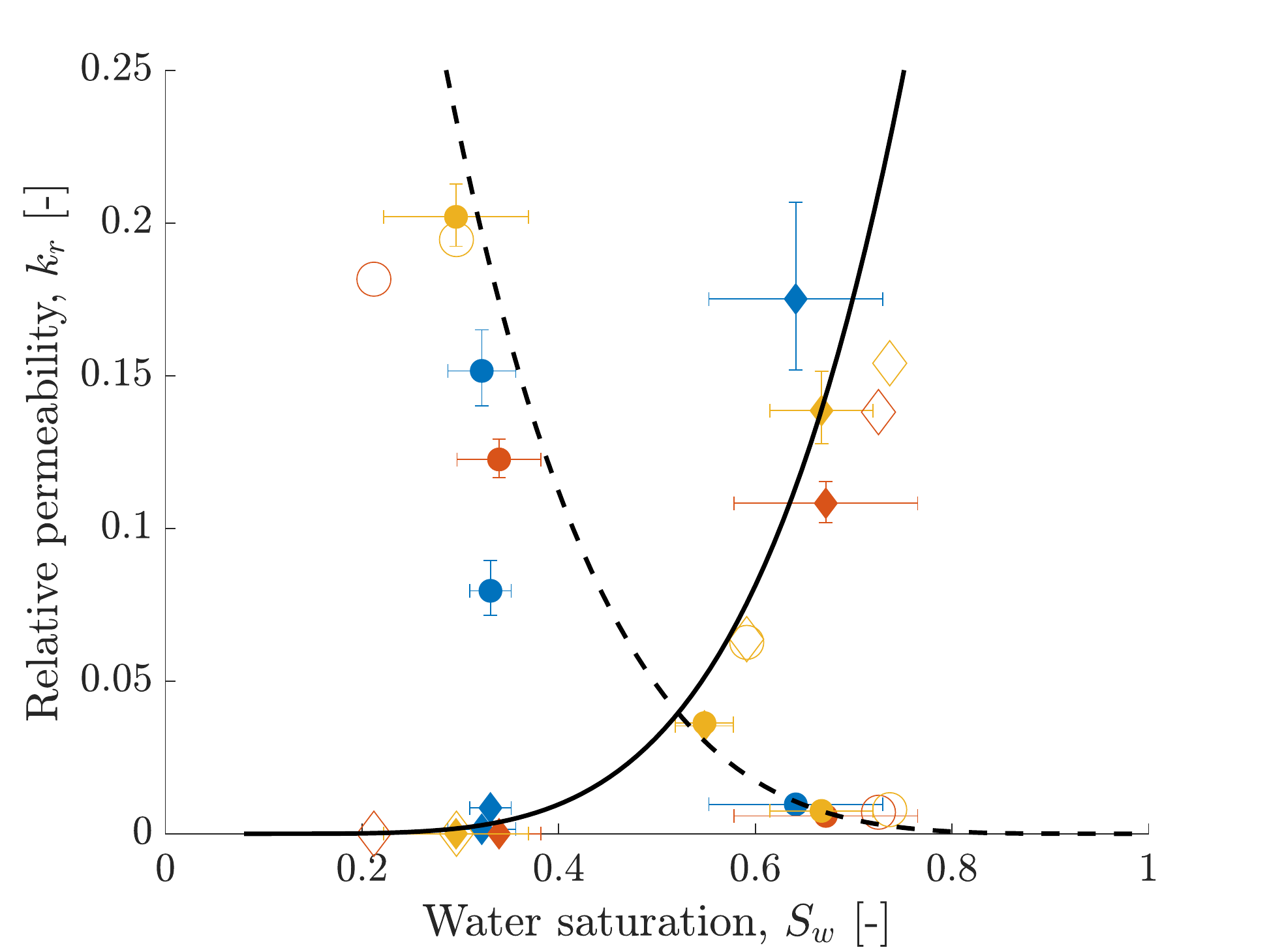}}
\subfloat[]{\includegraphics[width=0.45\textwidth]{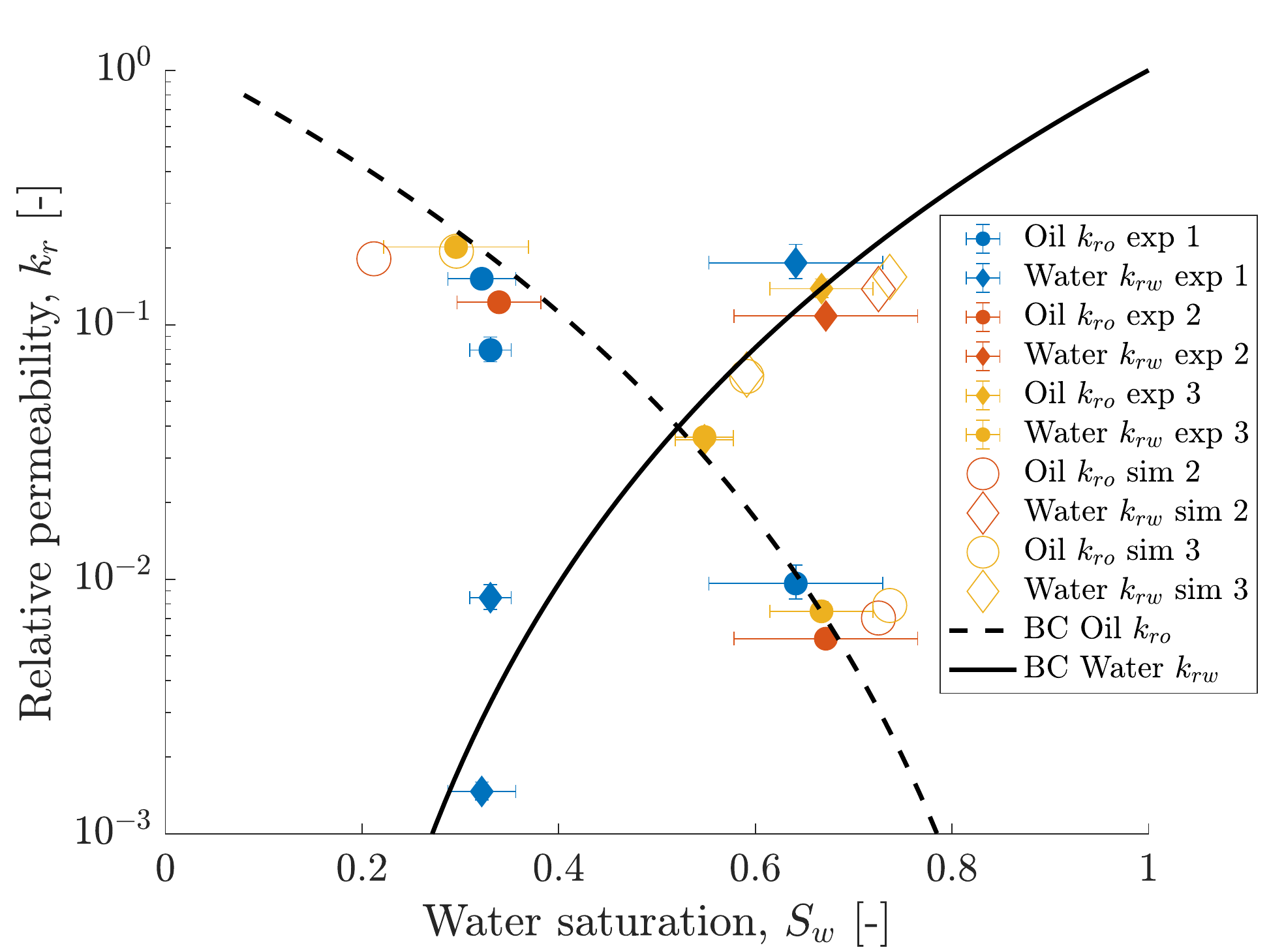}}

\caption{Whole core saturations and relative permeabilities for the experiments and simulations. Simulations are shown for the model derived from SR LT images, further realizations can be found in the SI. (a, b) Voxelised saturations for core 1 and 2, respectively. (c, d) Slice average saturations for core 1 and 2, respectively.  (e, f) Linear and logarithmic relative permeability results, respectively. Open symbols are the simulations, closed symbols are the experiment. Note, here we simulate exp 2 and exp 3 from \cite{Jackson2020}.}
\label{fig_whole_core_sat}
\end{figure*}

Further from the single-phase predictions, we present multiphase flow predictions in the form of relative permeability and saturation in Figure \ref{fig_whole_core_sat} for the SR LT model, and tabulated results for the LR and SR LT models in Table \ref{tab_multiphase} (further results are in the SI: Table S1 shows LR and SR HT results and Figures S10 - S12 show the LR LT, LR HT and SR HT model results, respectively). We see that core average relative permeabilities are generally well predicted by the SR LT model along with 3D voxel and slice average saturations along the profile of the core. Overall, the SR LT model represents the most accurate model when considering both saturation and pressure predictions. Core average saturations are predicted to within the experimental repeatability ($\pm 10\%$), apart from $f_o  = 1$ in core 1. This is highlighted in Figure \ref{fig_whole_core_sat}a,c whereby the saturations are correlated, but raised relative to the experimental results. There are strong capillary-end effects in the model, which are not as prominent in the experiment. Higher threshold models in Figures S10 - S12 show less end-effects, due to a lower $P_c$ transition at the outlet and may be more representative of the true experiment. In the models, we use a constant, core average capillary pressure at the outlet ($P_c = P_e = 3.7$kPa), and note that the model is slightly shorter than the full core due to imaging artefacts. The images are missing $\approx$ 5mm from each end, equivalent to 2.5 grid blocks (the missing regions are visible in the lack of data at $x/L < 0.07, x/L  > 0.93$).  Without calibration of the outlet $P_c$ \cite{Jackson2020} or artificial extensions to the model geometry \cite{Zahasky2020}, we generally overpredict end-effects at high $f_o$. We choose to have a calibration free model here, which provides robust results for cases with small end-effects ($f_o < 1$) and allows bias free modelling of the process. 

Generally, we see that the saturation predictions of the LR and SR models are similar, however the SR model gives a marginally better representation of saturation heterogeneity in the core, likely due to a slightly more accurate capillary pressure near the percolation threshold. The relatively good performance of both models suggests that saturation heterogeneity is not strongly linked to the functional representation of capillary pressure here, but instead related to the relative variation in capillary pressure through the core, as shown in \cite{Zahasky2020}. Even though the LR capillary pressures do not capture the overall shape from $0.2 < S_w < 0.8$ as well as the SR model (see Figure S8c, f and Figure S9c, f)) they still capture the relative variation between regions and hence saturation heterogeneity in the model. We see that relative permeability heterogeneity (see Figure S8a,b,d,e and Figure S9a,b,d,e) does not contribute strongly to saturation heterogeneity, also shown in \cite{Zahasky2020}. 

\begin{table*}
\caption{Multiphase flow simulation results compared to experiments. Results are shown for models derived from the low threshold (LT) images. Lines showing simulation error [\%] are in bold.The high threshold (HT) results are shown in the supporting information.}
\label{tab_multiphase}
\begin{tabular}{|l|c|c|c|c|c|c|c|c|c|c|}
\hline
\multirow{2}{*}{}                                    & \multicolumn{4}{c|}{\textbf{Core 1}}                                & \multicolumn{6}{c|}{\textbf{Core 2}}                                                          \\ \cline{2-11} 
                                                     & \multicolumn{2}{c|}{\textbf{LR}} & \multicolumn{2}{c|}{\textbf{SR}} & \multicolumn{3}{c|}{\textbf{LR}}              & \multicolumn{3}{c|}{\textbf{SR}}              \\ \hline
\textbf{Fractional   flow oil, $\bm{f_o}$   {[}-{]}} & 0.05           & 1               & 0.05           & 1               & 0.05          & 0.5           & 1             & 0.05          & 0.5           & 1             \\ \hline
$S_w$ av. exp {[}-{]}                                & 0.672          & 0.339           & 0.672          & 0.339           & 0.667         & 0.548         & 0.296         & 0.667         & 0.548         & 0.296         \\ \hline
$S_w$ av. sim {[}-{]}                                & 0.726          & 0.256           & 0.725          & 0.212           & 0.730         & 0.589         & 0.299         & 0.737         & 0.591         & 0.296         \\ \hline
\textbf{$\bm{S_w}$   av. err {[}\%{]}}               & \textbf{8.1}   & \textbf{-24.6}  & \textbf{8.0}   & \textbf{-37.5}  & \textbf{9.4}  & \textbf{7.4}  & \textbf{1.1}  & \textbf{10.4} & \textbf{7.8}  & \textbf{0.1}  \\ \hline
$k_{ro}$   exp {[}-{]}                               & 0.00584        & 0.123           & 0.00584        & 0.123           & 0.00748       & 0.0362        & 0.202         & 0.00748       & 0.0362        & 0.202         \\ \hline
$k_{ro}$ sim {[}-{]}                                 & 0.00784        & 0.238           & 0.00704        & 0.182           & 0.00783       & 0.0629        & 0.230         & 0.00789       & 0.0627        & 0.195         \\ \hline
\textbf{$\bm{k_{ro}}$ err {[}\%{]}}                  & \textbf{34.2}  & \textbf{94.4}   & \textbf{20.6}  & \textbf{48.2}   & \textbf{4.6}  & \textbf{73.6} & \textbf{13.8} & \textbf{5.4}  & \textbf{73.0} & \textbf{-3.7} \\ \hline
$k_{rw}$   exp {[}-{]}                               & 0.108          & 0               & 0.108          & 0               & 0.139         & 0.0353        & 0             & 0.139         & 0.0353        & 0             \\ \hline
$k_{rw}$ sim {[}-{]}                                 & 0.162          & 0               & 0.138          & 0               & 0.160         & 0.0664        & 0             & 0.154         & 0.0637        & 0             \\ \hline
\textbf{$\bm{k_{rw}}$ err {[}\%{]}}                  & \textbf{49.3}  & \textbf{0}      & \textbf{27.6}  & \textbf{0}      & \textbf{15.7} & \textbf{87.9} & \textbf{0}    & \textbf{11.2} & \textbf{80.3} & \textbf{0}    \\ \hline
\end{tabular}
\end{table*}

The relative permeability predictions from the SR LT model are more accurate than the LR LT model and the other HT models. We are able to capture the relative permeability to within the same order of magnitude (often $<20\%$ error) across three orders of magnitude in range. The non-wetting core average relative permeability (Figure Figure \ref{fig_whole_core_sat}e,f) in each cases is equivalent to the average of the individual subvolume relative permeabilities (Figure S8, S9) with some degree of end-effect depending on the fractional flow, with larger impacts at higher $f_o$. The type of average is likely harmonic, or similar, given the lower bounding of the subvolume relative permeabilities by the core-average results. For the wetting relative permeability, the core average looks to be a simple arithmetic average of the subvolume individual values (see Figure S8b,e, Figure S9b,e). This is physically intuitive since the wetting fluid remains connected at all modelled saturations.

Similar to the saturation prediction, the LR model prediction of relative permeability is worse than the SR, but not significantly, see Table \ref{tab_multiphase}. However, we note that the pressure predictions are significantly worse in the LR models during both single and multiphase flow compared to the SR models. The less accurate pressure prediction does not necessarily translate directly to relative permeability, since relative permeability essentially acts as a ratio between the core averaged single phase and multiphase permeability at a given saturation. In the low capillary number, high $S_w$ flow regime here, flow is controlled strongly by the largest percolating pathway through the core. The flow is well modelled through this pathway, even at low resolution, with the other regions behaving as single phase flow conduits. These single phase flow regions have the same inaccurate pressure contribution as per the absolute permeability calculation, and therefore essentially cancel out when considering the relative permeability. For more uniform displacement fronts, at higher capillary number, or at lower $S_w$ it is likely that the SR model will behave significantly better than the LR model as more of the pore-space is utilised and the resolution of smaller throats will become more important. This will also be true in systems with more heterogeneous flow paths and larger ratios between pore body to throat sizes (e.g., carbonate systems). 

We see that the relative permeability at $f_o = 0.5$ in the core 2 experiment is not well captured by any model, and suggests an epistemic uncertainty. In the models, we only simulate the drainage process, but during the $f_o = 0.5$ co-injection, it is likely that both drainage and imbibition displacement mechanisms are occurring. If local trapping is also occurring, this could explain the lower experimental $k_r$ due to an increased resistance to flow. It is also possible that flow intermittency \cite{Reynolds2017a} is occurring at this fractional flow due to the increased multiphase competition, which is not represented in the model. Intermittency generally results in more energy dissipation, meaning relative permeabilities would be lower than in the connected pathway case. 

Overall, we see that the SR models are able to accurately reproduce the experimental behaviour of the continuum-scale system, and are generally more accurate than the basic LR derived models. We can capture the key saturation and pressure behaviour over a range of conditions during drainage. For the experimental conditions simulated in this paper, the largest improvements from the SR model are seen in pressure and absolute permeability predictions, which have contributions from a wide range of pore body and throat sizes. We also see improvements in capturing saturation heterogeneity (see the voxel saturation plots) and relative permeability in the SR model, but the benefits are not as strong when flow is controlled by large, connected pore body and throats, which are already well resolved in the LR model. This is a consequence of the resolution chosen and the available experimental data, which constrain the `low' resolution data. Based on this, further exploration of lower resolution systems is warranted, perhaps moving between $20\mu$m and $1-2\mu$m resolutions (in multiple steps to aid information transferal) where the underlying full core images are themselves not able to accurately capture the main percolating pathways. 

\section{Conclusion}

In this work, we have developed an EDSR neural network to overcome traditional field-of-view and resolution trade-offs in X-ray micro-CT imaging of multi-scale porous media. Distinct from previous work, we developed the deep learning architecture in 3D using paired LR and HR data from optically magnified X-ray CT images across multiple subvolumes. We tested the network on unseen LR and HR data from the relatively homogeneous Bentheimer used for training, as well as a more heterogeneous Bentheimer sample with a distinct micro-structure; low permeability bands with significant hematite mineral inclusions. The proposed network architecture worked well at alleviating common imaging defects (ring, beam hardening) that are common in high-resolution micro-CT images. The 3D EDSR network was able to produce physically realistic SR images from the input LR images, that emulated the true HR images across a variety of image measures (SSIM, textural analysis) as well as in multiphase flow simulations using pore-network modelling. We provided a robust validation of the SR pore-scale behaviour using images from several segmentation realizations, removing image processing bias. 

We used the trained EDSR network to validate the approach against experimental data from \cite{Jackson2020}. We generated $\approx$1000 high-resolution REV scale subvolumes for each whole core from the low resolution data. With this, we populated a continuum model, at cm scale, with petrophysical properties generated from the subvolume pore-network models. The continuum model was used to simulate drainage immiscible multiphase flow at low capillary number across a range of fractional flows, comparing directly to experiments. We found the EDSR generated model was more accurate than the base LR model at predicting experimental behaviour in the presence of heterogeneities, especially in flow regimes where a wide distribution of pore-sizes were encountered. The models were generally accurate at predicting saturations to within the experimental repeatability and relative permeability across three orders of magnitude. We found less improvement in the EDSR model when the main flow paths were already captured by the LR images, with saturation heterogeneity and the (low capillary number) relative permeability largely controlled by the largest pores in the domain. 

The demonstrated methodology is fully deterministic and opens up the possibility to image, simulate, and analyse flow in truly multi-scale heterogeneous systems that are otherwise intractable. Further from the digital rock physics applications in this work, the methodology can be useful generally in materials science research with multi-scale porous systems. In particular, the design and optimisation of heterogeneous lithium-ion batteries could be improved with accurate multi-scale characterisation \cite{Lu2020}, and for hierarchical phase change systems with nano and macro-porous components \cite{Grosu2020}.

As well as the multi-scale aspects considered in this work, the methodology also has the potential to improve time-resolved X-Ray CT imaging, by permitting coarser and therefore faster imaging of large systems. When combined with recent deep-learning noise-removal techniques \cite{Niu2021} this could significantly speed up acquisition times and allow the imaging of pore-scale dynamic processes over continuum-scales in the laboratory.

Further work considering greater differences in resolution between LR and HR data is being considered using a multi-scale deep super resolution network as per \cite{Lim2017}. Here, training data is required at multiple resolutions, with information linked across scales. This could represent a viable tool for increasing scalability further. Alongside this, approaches that do not require paired input data, e.g., based on generative adversarial networks (GANs) \cite{Niu2020, Niu2021a} could be utilised. Initial results from GANs trained on un-paired data suggests they can achieve similar performance to a CNN trained on paired data \cite{Niu2021a}. This could open up the possibility of training with decoupled images from data storage platforms to exploit the wide range of available data already present in the literature. With this, unpaired data at different resolutions with a multi-scale network could also be exploited. 

The data and codes used in this work are open access, discussed in Section \ref{sec_data_access} below; which provide a robust dataset to test different deep-learning models. We welcome all future developments on this topic. 

\section{Acknowledgements}
\label{sec_acknowledgements}

We gratefully thank Dr. Samuel Krevor from Imperial College London for access to the Zeiss micro-CT scanner. We acknowledge Computer Modelling Group (CMG) for providing access to IMEX. The authors have no conflicts of interests.  

\section{Data access}
\label{sec_data_access}

Data associated with this work is available in two locations. The original dataset from \cite{Jackson2020} is hosted on the BGS National Geoscience Data Centre, ID $\#$130625 at \textcolor{blue}{\href{http://dx.doi.org/10.5285/5f899de8-4085-4370-a45e-e613f27e8f1d}{dx.doi.org/10.5285/5f899de8-4085-4370-a45e-e613f27e8f1d}} (there is also a subvolume image dataset, for easier download available on the Digital Rocks Portal, project 229, DOI:10.17612/KT0B-SZ28 at \textcolor{blue}{\href{http://digitalrocksportal.org/projects/229}{digitalrocksportal.org/projects/229}}). The new LR (6 $\mu$m resolution), HR (2 $\mu$m resolution) and additional LR (18 $\mu$m resolution) subvolume images are hosted on \textcolor{blue}{\href{https://doi.org/10.5281/zenodo.5542623}{zenodo.org}} with DOI: 10.5281/zenodo.5542624. We host the deep learning code, as well matlab files for processing the images, and generating input files for the PNM and continuum modelling on Github at  \textcolor{blue}{\href{http://github.com/sci-sjj/EDSRmodelling}{http://github.com/sci-sjj/EDSRmodelling}}. Users can download the pnextract and pnflow software from  \textcolor{blue}{\href{http://github.com/ImperialCollegeLondon/pnextract}{github.com/ImperialCollegeLondon/pnextract}} and  \textcolor{blue}{\href{http://github.com/ImperialCollegeLondon/pnflow}{github.com/ImperialCollegeLondon/pnflow}}. IMEX is available at cost from CMG, however equivalent results have been obtained for the continuum modelling approach using the open-source MOOSE package  \textcolor{blue}{\href{http://github.com/idaholab/moose}{github.com/idaholab/moose}} running the finch application at  \textcolor{blue}{\href{http://github.com/cpgr/finch}{github.com/cpgr/finch}}. Examples are provided therein for running core-flood simulations, and the CMG input files can be readily converted to the MOOSE syntax. 


%

\end{document}